\documentclass[preprint,sort&compress,12pt]{elsarticle}
\usepackage{amsmath}
\usepackage{mathrsfs}
\usepackage{stmaryrd}
\usepackage{mathrsfs}
\usepackage{bm}
\usepackage{amsmath}
\usepackage{amsfonts}
\usepackage{amssymb}
\usepackage{amsthm}

\newcommand{\mm}{\mathrm}

\newcommand{\dx}{\mathrm{d}x}

\begin{document}
\begin{frontmatter}
\title{On Linear Instability and Stability  of the Rayleigh-Taylor\\ Problem in Magnetohydrodynamics}

\author[FJ]{Fei Jiang\corref{cor1}
}
\ead{jiangfei0591@163.com}
 \cortext[cor1]{Corresponding
author.}
\author[sJ]{Song Jiang}
\ead{jiang@iapcm.ac.cn}
\address[FJ]{College of Mathematics and
Computer Science, Fuzhou University, Fuzhou, 350108, China.}
\address[sJ]{Institute of Applied Physics and Computational Mathematics,
 Beijing, 100088, China.}

\begin{abstract} We investigate the stabilizing effects of the magnetic fields in the linearized magnetic Rayleigh-Taylor (RT)
problem of a nonhomogeneous incompressible viscous magnetohydrodynamic fluid of zero resistivity in the presence of a uniform
gravitational field in a three-dimensional bounded domain, in which the velocity of the fluid is non-slip on the boundary.
By adapting a modified variational method and careful deriving \emph{a priori} estimates, we establish a criterion for the instability/stability
of the linearized problem around a magnetic RT equilibrium state. In the criterion, we find a new phenomenon that a sufficiently strong
horizontal magnetic field has the same stabilizing effect as that of the vertical magnetic field on growth of the magnetic RT
instability. In addition, we further study the corresponding compressible case, i.e., the
Parker (or magnetic buoyancy) problem, for which the strength of a horizontal magnetic field decreases with height,
and also show the stabilizing effect of a sufficiently large magnetic field.
 \end{abstract}

\end{frontmatter}


\newtheorem{thm}{Theorem}[section]
\newtheorem{lem}{Lemma}[section]
\newtheorem{pro}{Proposition}[section]
\newtheorem{cor}{Corollary}[section]
\newproof{pf}{Proof}
\newdefinition{rem}{Remark}[section]
\newtheorem{definition}{Definition}[section]
\section{Introduction}
\label{Intro} \numberwithin{equation}{section}

The three-dimensional (3D) nonhomogeneous, incompressible and viscous magnetohydrodynamic
(MHD) equations with zero resistivity (i.e. without magnetic diffusivity)
in the presence of a uniform gravitational field in a domain $\Omega
\subset \mathbb{R}^3$ reads as follows
(see, for example, \cite{CHTA1,cowling1957magnetohydrodynamics,kulikovskiy1965magnetohydrodynamics,
landau1984electrodynamics}
on the derivation of the equations):
\begin{equation}\label{0101}\left\{\begin{array}{l}
 \rho_t+\mm{div}(\rho{  v})=0,\\[1mm]
\rho v_t+\rho v\cdot\nabla v +\nabla( p+\lambda_0\nabla |M|^2/2)=\mu\Delta v
+ \lambda_0M\cdot \nabla M-\rho g e_3, \\[1mm]
M_t=M\cdot \nabla v-v\cdot\nabla M, \\[1mm]
\mathrm{div} {v}=\mathrm{div} M=0 .\end{array}\right.\end{equation}
Here the unknowns $\rho =\rho(x,t)$, ${v}= {v}(x,t)$,
$M:= {M}(x,t)$ and $p=p(x,t)$ denote the density, velocity,
magnetic field and pressure of the incompressible fluid, respectively; $\mu>0$ stands for the coefficient
of shear viscosity, $\lambda_0$ for the permeability of vacuum, $g>0$ for the gravitational constant,
$e_3=(0,0,1)^{\mm{T}}$ for the vertical unit vector, and $-\rho g {e}_3$ for the gravitational force.
In the system \eqref{0101} the equation \eqref{0101}$_1$ is the continuity equation,
\eqref{0101}$_2$ describes the balance law of momentum, while \eqref{0101}$_3$ is called the induction equation.
As for the constraint $\mm{div}\, M=0$, it can be seen just as a restriction on the initial value of $M$ since
$(\mm{div}\,M)_t=0$ due to (1.1)$_3$. We remark that the resistivity is neglected in \eqref{0101}$_3$,
 and this arises in the physics regime with negligible electrical resistance.

Now, we choose a Rayleigh-Taylor (RT) density profile $\bar{\rho}:=\bar{\rho}(x_3)$, which is independent of $(x_1,x_2)$ and satisfies
\begin{eqnarray}\label{0102}
\bar{\rho}\in C^{1}(\bar{\Omega}),\quad\;\inf_{ x\in \Omega}\bar{\rho}>0,\quad\;\bar{\rho}'|_{x_3=x^0_{3}}>0\;
\mbox{ for some }x^0_{3}\in \{x_3~|~(x_1,x_2,x_3)\in \Omega\},
\end{eqnarray}
where $\bar{\rho}':=\mm{d}\bar{\rho}/\mm{d}x_3$, $x_{3}^0$ is the third component of $x_0\in \Omega$.
We refer to \cite[Remark 1.1]{NJTSC2} for the construction of such $\bar{\rho}$.
We remark that the second condition in \eqref{0102} prevents us from treating vacuum in the construction of unstable solutions,
while the third one in \eqref{0102}
assures that there is at least a region in which the RT density has larger density with increasing height $x_3$ and may lead to
the classical RT instability. Since we investigate the stabilizing effects of the horizontal and vertical magnetic fields,
we consider the magnetic field profile $\bar{ {M}}=me_1$ (i.e. the magnetic field orthogonal
to the direction of the gravitational force) or $me_3$  (i.e., the magnetic field parallel to the direction of the gravitational force),
where $e_1:=(1,0,0)^{\mm{T}}$ and $m$ is non-zero constant. Then $(\rho ,v, M)=(\bar{\rho},0,\bar{M})$ defines a magnetic RT
equilibrium state to \eqref{0101}, where the equilibrium pressure profile $\bar{p}$ is determined by
\begin{equation*}
\nabla \bar{p}=-\bar{\rho}g  {e}_3.
\end{equation*}

Denoting the perturbation to the RT equilibrium state by
$$ \varrho=\rho -\bar{\rho},\quad  {u}= {v}- {0},\quad N= M -\bar{M},\quad q=p-\bar{p},$$
then, $(\varrho , {u},q)$ satisfies the perturbed equations
\begin{equation}\label{0103}
\left\{\begin{array}{l}  \varrho_t+{  u}\cdot\nabla (\varrho+\bar{\rho})=0, \\[1mm]
(\varrho+\bar{\rho}){  u}_t+(\varrho+\bar{\rho}){  u}\cdot\nabla
{  u}+\nabla (q+\lambda_0 | N+\bar{M} |^2/2) \\
\quad\; =\mu \Delta u + \lambda_0(N+\bar{M})\cdot \nabla (N+\bar{M})-(\varrho+\bar{\rho}) g e_3,\\[1mm]
N_t=(N+\bar{M})\cdot \nabla u- u\cdot\nabla (N+\bar{M}),\\[1mm]
\mathrm{div} {u}= \mathrm{div} {N}=0.\end{array}\right.  \end{equation}
For the system (\ref{0103}), we impose the initial and boundary conditions:
\begin{eqnarray} &&  \label{0104}
(\varrho ,u,N)|_{t=0}=(\varrho_0,u_0,N_0)\quad\mbox{in } \Omega ,\\[1mm]
&&  \label{0105} u(x,t)|_{\partial\Omega}={  0}\quad \mbox{ for any }t>0.
\end{eqnarray}
If the perturbation $(\varrho,u,N)$ is very small, then the
two-order small terms (i.e., the nonlinear terms) could be neglected in the perturbed equations, and
we obtain the following linearized magnetic RT equations around the equilibrium state $(\bar{\rho},0,\bar{M})$:
\begin{equation}\label{0106}
\left\{\begin{array}{ll}
 \varrho_t+\bar{\rho}'{u}_3=0, \\[1mm]
  \bar{\rho} u_t +\nabla (q+\lambda_0 m N_i)=\mu\Delta u +\lambda_0 m\partial_i N -\varrho ge_3,\\[1mm]
   N_t=m\partial_i u, \\[1mm]
 \mathrm{div} u =\mathrm{div} N =0, \qquad\qquad i=1,3.
\end{array}\right.\end{equation}

The linearized equations are convenient in mathematical analysis in order to have an insight into the physical
and mathematical mechanisms of the magnetic RT instability. In this article, we investigate the instability/stability
of the linearized magnetic RT problem \eqref{0104}--\eqref{0106} in a bounded domain.

Next we briefly introduce the related background and research motivation of the magnetic RT instability.
The RT instability is well-known as a gravity-driven instability in fluid dynamics when a heavy fluid is on top
of a light one. This phenomenon was first studied by Rayleigh 1883 \cite{RLAP} and then Taylor, thus called Rayleigh-Taylor instability.
The analogue of the RT instability arises when fluids are electrically conducting and a magnetic
field is present, and the growth of the instability will be influenced by the magnetic field due to the generated
electromagnetic induction and the Lorentz force. In the last decades, the stabilizing effect of the magnetic field
on RT instability has been analyzed by a number of authors.
Kruskal and Schwarzchild in 1954 first showed that a horizontal magnetic field has no effect on growth of
the linear RT instability \cite{KMSMSP}, and such RT instability arising in magnetohydrodynamics is called
the Kruskal-Schwarzschild instability. Then the stabilizing role of a vertical magnetic field was further investigated
by Hide in \cite{HRWP} where the effect of finite viscosity and resistivity was included and his analysis
was encumbered with many parameters. These results were summarized in the monograph \cite{CS} by Chandrasekhar.
Similar results also hold for the problem \eqref{0104}--\eqref{0106} defined in a horizontally
periodic domain (i.e., $\Omega:= (2\pi L\mathbb{T})^2\times (-l,l)$,
where $0<l\leq +\infty$ and $2\pi L \mathbb{T}$ stands for the 1D-torus of length $2\pi L$).
More precisely, one can construct an unstable solution to \eqref{0104}--\eqref{0106}
for any horizontal magnetic field $\bar{M}=me_1$ and for the vertical magnetic field $\bar{M}=me_3$
with $m<M_{\mm{C}}$, where $M_{\mm{C}}$ denotes a critical number defined by
 \begin{equation}\label{origcrictical}
M_\mathrm{C}:=\sqrt{\sup_{\psi\in H^1_0(-l,l) }
\frac{g\int_{-l}^l\bar{\rho}'|\psi(x_3)|^2\mm{d}x_3}{\lambda_0 \int_{-l}^l|\psi'(x_3)|^2\mm{d}x_3}}>0,
 \end{equation}
 see \cite[Theorem 1.1]{JFJSWWWN} for more details. Moreover, if $l\in (0,+\infty)$, one can show
 the stability of \eqref{0104}--\eqref{0106} for any $m>M_{\mm{C}}$, please refer to \cite{WYC}
 for the proof of the linear stability. This means that a sufficiently strong vertical magnetic field
 can prevent the growth of the linear RT instability. These different effects on instability
 between the horizonal and vertical magnetic fields motivate us to study the intrinsic mechanism from the
 mathematical point of view. By a simple analysis of the Lorentz force, we find that
 the non-slip boundary condition of the velocity can play an important role in the stabilizing effect
 of the vertical magnetic field. On the other hand, we also notice that the velocity is horizontally periodic
 and non-slip at $x_3=\pm l$ for $l\in (0,+\infty)$, i.e, the boundary conditions for the
velocity are different in the horizontal and vertical directions.
This shows that the different behaviors on instability of the horizonal and vertical magnetic fields
may be somehow related to the boundary condition of the velocity. The same situation occurs for
the inviscid magnetic RT problem \cite{HHVQ} (i.e. $\mu=0$ in  \eqref{0105}--\eqref{0106})
in a 2D periodic domain, where the velocity is periodic in both vertical and horizontal directions.
 A natural question arises whether the horizontal magnetic field has the same stabilizing effect
 as the vertical one on growth of the magnetic RT instability, when the velocity is non-slip
 on the boundary of a bounded domain $\Omega$.

   The first aim of this article is to give a positive answer, namely, we shall find that there is a critical number
   $m_{\mm{C}}^{\mm{3}}$ for a horizontal magnetic field $\bar{M}=m e_3$, such that the linearized
 magnetic RT steady state is stable provided $m>m_{\mm{C}}^{\mm{3}}$ (see Theorem \ref{thm:0201}).
 Moreover, we find that the horizontal magnetic field has the same stabilizing effect as the vertical one
 on growth of the magnetic RT instability (see Remark \ref{rem:0203}).
 Thus, this article updates Kruskal and Schwarzschild's results.
 To our best knowledge, this is the first article to study the role of the non-slip boundary condition of velocity in the horizontal direction in the MHD instability under a horizontal magnetic field. We also mention that the magnetic RT instability for two-layer
incompressible fluids separated by a free interface (stratified fluids) in a horizontally periodic domain is investigated and
some weakly nonlinear instability results are obtained, see \cite{JFJSWWWOA}.

The second aim is to extend Theorem \ref{thm:0201} for the incompressible MHD problem \eqref{0105}--\eqref{0106} with $\bar{M}=me_1$
to the corresponding compressible isentropic case, in which the horizontal magnetic field is vertically stratified.  Before stating our
result, we introduce the governing equations. The corresponding compressible isentropic model of \eqref{0101} reads as follows.
\begin{equation}\label{comequations}\left\{\begin{array}{l}
 \rho_t+\mm{div}(\rho{  v})=0,\\[1mm]
\rho v_t+\rho v\cdot\nabla v+\nabla
   (p+  \lambda_0 |M|^2/2) =\mu\Delta v+\mu_0\nabla\mm{div} v
+\lambda_0M\cdot \nabla M-\rho g e_3,\\[1mm]
M_t=M\cdot \nabla v-v\cdot\nabla M - M \mm{div }v ,\\[1mm]
\mathrm{div} {M}=0,\end{array}\right.\end{equation}
where $\mu_0:=\mu+\nu$, $\nu$ denotes the bulk viscosity and $3\nu+2\mu\geq 0$. The pressure $p$ is
usually determined through the equations of state. In this article we focus our study on the
isentropic flow case
and consider that
\begin{equation}\label{isetropiceest}
p\equiv p(\rho)=A\rho^\gamma,
\end{equation}
where $\gamma\geq 1$ denotes the adiabatic constant and $A>0$ is a constant.

Next we construct a magnetic RT equilibrium state for \eqref{comequations}.
Letting $\bar{\rho}$ satisfy \eqref{0102}, we define a horizontal magnetic field profile $\bar{M}_{\mm{c}}:= (m_{\mm{c}},0,0)$ with
\begin{equation}\label{com:01091}
   m_{\mm{c}}  =\pm\sqrt{\frac{2}{\lambda_0}\left(C
- p(\bar{\rho}) -gF(\bar{\rho})\right)}, \qquad(\mbox{a function of }x_3)
\end{equation}
where $F(\bar{\rho})$ denotes a primitive function $\bar{\rho}$ and $C$ is a positive constant satisfying
$$ C - p(\bar{\rho}) -gF(\bar{\rho})>0 \quad\mbox{ on }\bar{\Omega}.$$
It is easy to see that
  \begin{equation}\label{comsteady}
  p'(\bar{\rho})\bar{\rho}'=-\lambda_0    m_{\mm{c}} m'_{\mm{c}} -g\bar{\rho},
 \end{equation}
 where $p'(\bar{\rho}):=A\gamma\bar{\rho}^{\gamma-1}$. Thus $(\bar{\rho},0,\bar{M}_{\mm{c}})$
 constructed above is an equilibrium state  to \eqref{comequations}, i.e.,
\begin{equation}\label{equcomre}
 \nabla  (p(\bar{\rho})+\lambda_0  |\bar{M}_{\mm{c}}|^2/2)= \lambda_0\bar{M}_{\mm{c}}\cdot \nabla \bar{M}_{\mm{c}}
 -\bar{\rho} g e_3\quad\mbox{and}\quad\mathrm{div} \bar{M}_{\mm{c}}=0.  \end{equation}
Obviously, by virtue of the relation \eqref{comsteady}, $m_{\mm{c}}$ is impossible to be a constant.
We remark here that, by virtue of the relation \eqref{equcomre}, there does not exist a magnetic RT equilibrium state $(\bar{\rho},0, \bar{M})$
 of \eqref{comequations}, such that $\bar{M}$ is a vertical magnetic field.

 Now denoting the perturbation to $(\bar{\rho},0,\bar{M}_{\mm{c}})$ by
$$ \varrho=\rho -\bar{\rho},\quad  u= v- {0},\quad N=M-\bar{M}_{\mm{c}},$$
we get the perturbed equations:
\begin{equation} \left\{\begin{array}{l}
\varrho_t+\mm{div}((\varrho+\bar{\rho}){ u})=0, \\[1mm]
(\varrho+\bar{\rho}){  u}_t+(\varrho+\bar{\rho}){  u}\cdot\nabla
{  u}+\nabla (p(\varrho+\bar{\rho})+\lambda_0|N+\bar{M}_{\mm{c}}|^2/2)\\
\quad =\mu\Delta u+\mu_0\nabla\mm{div} u+
\lambda_0( N+\bar{M}_{\mm{c}})\cdot \nabla (N+\bar{M}_{\mm{c}})-(\varrho+\bar{\rho}) g e_3,\\[1mm]
 N_t=( N+\bar{M}_{\mm{c}})\cdot \nabla u
 -u\cdot\nabla (N+\bar{M}_{\mm{c}})-(N+\bar{M}_{\mm{c}})\mm{div}u,\\[1mm]
 \mathrm{div} N=0.\end{array}\right.    \label{js1} \end{equation}
We impose the following initial and boundary conditions for (\ref{js1}):
\begin{eqnarray} \label{c0104}   &&
(\varrho,{  u}, {N} )|_{t=0}=(\varrho_0,u_0,N_0)\;\quad\mbox{in } \Omega ,  \\
&& \label{c0105}
u(x,t)|_{\partial\Omega}=0 \;\quad \mbox{ for any }\; t>0.
\end{eqnarray}

The linearized equations of (\ref{js1}) around the equilibrium state $(\bar{\rho}, {0},\bar{ {M}})$ read as
\begin{equation}\label{lincom}
\left\{\begin{array}{ll}
 \varrho_t+\mm{div}( \bar{\rho} {u})=0, \\[1mm]
  \bar{\rho} u_t +\nabla ( p'(\bar{\rho})\varrho+ \lambda_0
 m_{\mm{c}} N_1)=\mu\Delta{ u}+\mu_0\nabla\mm{div}{u} + \lambda_0 N_3  \bar{M}'_{\mm{c}}+
  \lambda_0 m_{\mm{c}} \partial_1  N-\varrho  ge_3,\\[1mm]
   N_t=m_{\mm{c}} \partial_1 u -u_3 \bar{M}_{\mm{c}}'-\bar{M}_{\mm{c}}\mm{div}u,\\[1mm]
 \mathrm{div} N=0 ,   \end{array}\right.\end{equation}
where $M'_{\mm{c}}:=\mm{d}M_{\mm{c}}/\mm{d}x_3$. The system (\ref{lincom}) with suitable
initial and boundary conditions constitutes a compressible (viscous) magnetic RT problem.

Now, we introduce the related research  background on the compressible magnetic RT problem.
By virtue of the conditions \eqref{0102}, \eqref{isetropiceest} and \eqref{comsteady},
there exists at least a region in which the RT density profile has larger density with increasing height and the magnetic field
$\bar{M}_{\mm{c}}$ causes a non-zero Lorentz force  $ \lambda_0(\bar{M}_{\mm{c}}\cdot \nabla \bar{M}_{\mm{c}}  -\nabla |\bar{M}_{\mm{c}}|^2/2)$,
the direction of which is opposite to gravity, to support the heavier gas layered on top of lighter one. In particular, if the strength of
the Lorentz force increases in \eqref{equcomre}, we see that the Lorentz force plays a role of buoyancy to drive the heavier gas to go up.
This idea of ``magnetic buoyancy'' was introduced by Parker \cite{ParkerEN} in connection with the formation of sunspots. When sunspots
break out, most solar prominences will go up due to the magnetic buoyancy, and then slowly fall down toward  the surface of the sun due to
gravity of the sun, while some solar prominences can float in the solar corona for a long time, although the density of the former
is heavier 1000-10000 times than the latter. Hence, if the equilibrium state \eqref{equcomre} is slightly disturbed, then the instability
caused by magnetic buoyancy and gravity may occur in compressible MHD problems. Therefore, such an instability is commonly referred as
 the magnetic buoyancy instability due to the new characteristic caused by the magnetic buoyancy, also
 called the Parker instability in the astronomical literature \cite{ParkerEN1955}.
Thus, the linearized compressible magnetic RT problem \eqref{c0104}--\eqref{lincom} can be  called
the linearized Parker problem, we refer the reader to \cite{IADJSP23,HDGWAGTDX} and the references cited therein for more physical
background on the Parker instability. The Parker instability is also suggested to be responsible for other observed astrophysical
effects, for examples, Parker \cite{ParkerEN1955} demonstrated that the interstellar medium is unstable due to the magnetic buoyancy
as an instability mechanism, and thought that the Parker instability is associated with interstellar cloud formation.
It is worth pointing out that Fukui, et al. \cite{MLGFYS3}
observed giant molecular loops in a Galactic center, which offers a evidence for magnetic floating and supports thus Parker's theory.

The linear Parker instability have been widely investigated by physicists from the physical and numerical simulation points of view,
 see \cite{machida2013dynamo,kudoh2014magnetohydrodynamic,MKADCRHDW,barker2012magnetic,kuwabara2004nonlinear}
 and the references cited therein for example. In this paper, we study the  dynamical instability and stability of the linearized Parker problem
\eqref{c0104}--\eqref{lincom} from the mathematical point of view. More precisely, we give sufficiently conditions for the
linear instability and stability of the problem \eqref{c0104}--\eqref{lincom} in the Hadamard sense in Sobolev spaces.
Moreover, from the stability condition, we can easily see that a sufficiently strong $\bar{M}_{\mm{c}}$
has a remarkable stabilizing effect in the development of the Parker instability.
Therefore, our results for the compressible can be regarded as a generalization of those for the previous incompressible case.

The rest of this paper is organized as follows. In Section \ref{sec:02}
we introduce the instability and stability of both linearized magnetic RT and Parker problems.
Sections \ref{sec:03} and \ref{sect:04} are devoted to the proof of instability and stability of the two linearized problems. Finally,
we give an additional result on the critical number in a horizontally periodic domain and
prove the sharp growth rate of solutions to the two linearized problems in Section \ref{sec:05}.

\section{Main results}\label{sec:02}
 Before stating the main results, we introduce the notations used throughout this paper.
 We always assume that $\Omega$ be a $C^{2}$-smooth bounded domain. For simplicity, we
drop the domain $\Omega$ in Sobolev spaces and the corresponding norms as well as in integrands over $\Omega$, for example,
\begin{equation*}  \begin{aligned}&
L^2:=L^2(\Omega),\
{H}^1_0:=W^{1,2}_0(\Omega),\
{H}^k:=W^{k,2}(\Omega ),\;\;\mbox{ and } \int:=\int_\Omega. \end{aligned}\end{equation*}
We denote
\begin{equation*}  \begin{aligned}&
\mathbb{R}^+:=(0,\infty),\quad
{H}_{\sigma}^1:=\{{ {u}}\in {H}^1_0~|~
\mm{div}{ {u}}=0\},\quad J(w):=\int\bar{\rho}w^2\mm{d} x,\\
&\mathcal{A}=\left\{w\in H^1_0~\big|~J(w) =1\right\},\quad \mathcal{A}_\sigma=\mathcal{A}\cap  {H}^1_\sigma,\\
& f^0_j\mbox{ denotes the } j\mbox{-th component of the vector function } f_0.
\end{aligned}\end{equation*}
The letter  $C$ denotes a generic positive constant which may depend on $\Omega$ and other known physical quantities such as $g$,
$\bar{\rho}$, $\lambda_0$, $p$, $m$ and $m_{\mm{c}}$, but is independent of $\mu$ and $\mu_0$. Similarly, we denote by
$C_\mu$ still a generic positive constant to address the dependence on $\mu$.
In addition, a product space $(X)^n$ of vector functions is
still denoted by $X$, for example, a vector function $ {u}\in (H^2)^3$ is denoted
by $ {u}\in H^2$ with norm $\| {u}\|_{H^2}:=(\sum_{k=1}^3\|u_k\|_{H^2}^2)^{1/2}$.

\subsection{Incompressible case}
Our first main result is concerned with the instability/stability for the linearized incompressible magnetic
RT problem \eqref{0104}--\eqref{0106} and reads as follows.
\begin{thm}\label{thm:0201}
Denote the critical number by
\begin{equation}\label{criticalnumber}
m_{\mm{C}}^{i}:=\sup_{ { w}\in H_{\sigma}^1} \sqrt{\frac{g\int\bar{\rho}' { w}_3^2\dx}
 {\lambda_0\int|\partial_i  { w} |^2\dx}}\quad\qquad (i=1\mbox{ or }3).
 \end{equation}
 Assume that the density profile $\bar{\rho}$ satisfies \eqref{0102}.
Then, $m_{\mm{C}}^{i}$ is a threshold of $\bar M$ for instability and stability of the
problem \eqref{0104}--\eqref{0106} in the following sense:
\begin{itemize}[(1)]
  \item[(1)] If $|m|<m_{\mm{C}}^{ {i}}$, then the equilibrium state $(\bar{\rho},0,\bar{M})$ of the problem  \eqref{0104}--\eqref{0106}
 is unstable. That is, there is an unstable solution
$$( {\varrho}, {u},N,{q}):=e^{\Lambda t}(-\bar{\rho}'\tilde{u}_3/\Lambda,\tilde{ {u}},m\partial_i\tilde{u},\tilde{q})$$
 to  \eqref{0104}--\eqref{0106}, where $(\tilde{ {u}},\tilde{q})\in (H^2\cap \mathcal{A}_\sigma)\times H^1$
 solves the boundary value problem:
 \begin{equation}\label{0201}
\left\{    \begin{array}{l}
   \Lambda^2\bar{\rho}\tilde{u} =\Lambda\mu\Delta\tilde{ {{u}}}
-\nabla(\Lambda \tilde{q}+\lambda_0m^2\partial_i \tilde{u}_i) +    \lambda_0 m^2
 \partial_i^2 \tilde{u} +g\bar{\rho}'\tilde{u}_3e_3 , \;\;  \mm{div}\tilde{u}=0, \\
   \tilde{u}|_{\partial\Omega}= {0}  \end{array}  \right.
                       \end{equation}
  with a finite growth rate $\Lambda>0$ satisfying
\begin{equation}\label{Lambdard}
\begin{aligned}
\Lambda^2=\sup_{{ {w}}\in \mathcal{A}_\sigma} {\mathcal{E}_\sigma(w,\Lambda)}=
{\mathcal{E}_\sigma(\tilde{u},\Lambda)},  \end{aligned}\end{equation}
where
$$\mathcal{E}_\sigma(w,\Lambda):=\int g\bar{\rho}'w_3^2\mm{d} x-\lambda_0m^2\int|\partial_iw|^2\dx-\Lambda\mu\int|\nabla w|^2\mm{d} x.$$
Moreover,
\begin{equation}\label{0204n}
\begin{aligned}
&\| {\varrho}(t)\|_{L^2},\  \|(u_1,u_2)(t)\|_{L^2},\ \|u_3(t)\|_{L^2},\\
 &\| (N_1,N_2)(t)\|_{L^2}\;\;\mbox{ and }\;\;\| N_3(t)\|_{L^2}\to\infty\; \mbox{ as }t\to \infty.
\end{aligned}\end{equation}
  \item[(2)]If $|m|>m_{\mm{C}}^{ {i}}$ and the initial data $\varrho_0,u_0, N_0$ satisfy
 $$u_0\in H^1_\sigma\cap H^2,\  (\varrho_0,N_0,\partial_i N_0)\in L^2 \;\mbox{ and }\; \mm{div}N_0=0,$$
then there is a unique global solution $(\varrho,u, N)\in C^0(\mathbb{R}^+,L^2)\times C^0(\mathbb{R}^+ ,H^2)
\times C^0(\mathbb{R}^+,L^2)$ to the problem \eqref{0104}--\eqref{0106} satisfying the following stability estimates:
for any $t>0$,
 \begin{align} \label{uestimeate}
&\| (\varrho, {u}, N)_t(t) \|^2_{L^2}+\| ({u},\partial_i u)(t) \|^2_{L^2}+ \mu \int_0^t \| {u}_\tau\|^2_{H^1}\mm{d}\tau
\leq  C\|(\varrho_0,\partial_i{u}_0,\mu\Delta {u}_0,\partial_i N_0)\|_{L^2}^2,\\  \label{uestimeate2}
& \| u (t)\|^2_{L^2}+\left\|  \int_0^t (u, \partial_i u)(\tau)\mm{d}\tau \right\|^2_{L^2} +\mu \int_0^t \| {u}(\tau) \|^2_{H^1}\mm{d}\tau
\leq  C\| ( {u}_0, Q_0)\|_{L^2}^2,  \\[1mm]
\label{heighesimte}
 &\| (\varrho, N)(t) \|^2_{L^2} \leq  C\| (\varrho_0,  {u}_0, N_0,Q_0)\|_{L^2}^2,  \\[2mm]
\label{0324}
  & \mu\|\nabla{ u}(t)\|_{L^2}^2\leq C\| (\varrho_0, \partial_i{u}_0,\mu\Delta {u}_0, N_0,\partial_i N_0)\|_{L^2}^2,
\end{align}
where $Q_0:=\lambda_0 m\partial_{i} N_0-\varrho_0 ge_3$. Moreover, there exists a couple $(\varrho_\infty, N_\infty)\in L^2$, such that
\begin{align}
& \label{steadystate}\int (\lambda_0 m N_\infty \cdot \partial_{i}\varphi +g \varrho_\infty \varphi_3)\dx
=0\quad\mbox{for any }\varphi=(\varphi_1,\varphi_2,\varphi_3)\in H_\sigma^1, \\
\label{instabi}
&\|\varrho(t)-\varrho_\infty\|_{L^2},\; \| {u}(t)\|_{H^1},\;\| {u}_t(t)\|_{L^2}, \; \|N(t)-N_\infty\|_{L^2}\to 0
\;\; \mbox{ as }t\to\infty.  \end{align}
\end{itemize}
\end{thm}
\begin{rem}\label{rem:0201}
It should be remarked that we have omitted to write the restriction ${w}\neq 0$ in \eqref{criticalnumber} in order
to make the denominator sense. We mention the definition \eqref{criticalnumber} makes sense, and $m^i_{\mm{C}}\in (0,+\infty)$.
In fact, recalling the condition \eqref{0102} for $\bar{\rho}$, there exists a function $w\in H_\sigma^1$, such that
\begin{equation*}
 \int\bar{\rho}'{w}_{3}^2 \mm{d} {x}>0,
\end{equation*}
which guarantees the validity of \eqref{criticalnumber} and implies $m^i_{\mm{C}} >0$. On the other hand, we have
\begin{equation}\label{poincareinequality}
\|w_i\|_{L^2}\leq c_\Omega\|\partial_j w_i\|_{L^2}\quad\;\;\mbox{ for }\; w\in H^1_0,
\end{equation}
which implies $m^i_{\mm{C}}<+\infty$.
Here $c_\Omega$ is a positive constant depending on the diameter of $\Omega$ only.
\end{rem}
 \begin{rem}\label{rem:0203}
 For the linearized nonhomogeneous incompressible viscous fluid problem, it is shown in \cite[Theorem 1.1]{JFJSO2014}
 that a RT equilibrium state is always unstable. However, Theorem \ref{thm:0201} shows that in the presence of
 a sufficiently large magnetic field $\bar{M}$, the linear stability can be maintained due to
 the stabilizing effect of the magnetic field. In addition, when the domain is symmetric about
 the plane $x_1=x_3$ (e.g., a ball centered at the original point)  and $\bar{\rho}' $ is a constant, then
  \begin{equation*}\label{smaemsge}m_{\mm{C}}^{i} =\frac{g\bar{\rho}'}{\lambda_0}\sup_{ { w}\in H_{\sigma}^1}\sqrt{\frac{\int { w}_3^2\dx}
 {\int|\partial_i { w} |^2\dx}}\qquad\quad (i=1,3),
 \end{equation*}
and obviously, $m_{\mm{C}}^{1}=m_{\mm{C}}^{3}$.  This shows that
the horizontal magnetic field has the same stabilizing effect as the vertical one, a new phenomenon found in this article.
 \end{rem}
\begin{rem}\label{rem0202}
 Theorem \ref{thm:0201} still holds for a horizontally periodic domain with finite height (i.e.,
 $ \Omega:=(2\pi L\mathbb{T})^2\times (-l,l)$, $l\in (0,+\infty)$).
 In this case, we can assert that
   \begin{equation}\label{assertirem0202}
   \begin{aligned} m_{\mm{C}}^3=M_\mathrm{C},   \end{aligned}\end{equation}
   which will be shown in Section \ref{sec:05} directly by using the definitions \eqref{criticalnumber} and \eqref{origcrictical}.
In addition, we can see from  \eqref{criticalnumber} that $m_{\mm{C}}^1 = +\infty$ for the horizontally periodic domain. This means that
the horizonal magnetic field has no stabilizing effect (due to the horizontally periodic boundary condition).
\end{rem}
 \begin{rem}\label{remark0204}   We briefly describe the idea in the derivation of \eqref{0204n}.
 It follows from \eqref{0201} that $\tilde{u}\in H^1_0\cap H^2$ satisfies $\Lambda^2=\mathcal{E}_\sigma(\tilde{u},\Lambda)>0$;
 moreover, $\tilde{u}\in C^{0,\varsigma}(\Omega)$ for some constant $\varsigma\in (0,1)$ by Sobolev's embedding theorem.
 Thus, one has $\bar{\rho}'\tilde{{u}}_3 \not= 0$, $\tilde{{u}}_3 \not= 0$ and $\partial_i\tilde{u}_3 \not= 0$, and consequently,
$ \| {\varrho}(t)\|_{L^2}$, $\| u_3(t)\|_{L^2}$, $\| N_3(t)\|_{L^2} \rightarrow  \infty$ as $t\rightarrow \infty$.
Recalling that $\mm{div}\tilde{ {u}}=0$ and $\tilde{ {u}}|_{\partial\Omega}= {0}$, we get $\tilde{{u}}_1^2+\tilde{{u}}_2^2\not= 0$ and $|\partial_i\tilde{{u}}_1|^2+|\partial_i\tilde{{u}}_2|^2\not= 0$ immediately,
which yields $\|(u_1, u_2)(t)\|_{L^2}$ and $\| (N_1, N_2)(t)\|_{L^2}\rightarrow \infty$ as $t\rightarrow \infty$.
Such process is consistent with the phenomenon of the magnetic RT instability: gravity first drives the $3$-th
component of velocity unstable. Then, the instability of the $3$-th component of the velocity further results in instability of the density and
horizontal velocity. Finally, the instability of the velocity leads to instability of the magnetic field through the induction equation.
\end{rem}
\begin{rem}
 To our best knowledge, it is still open whether there exists a small smooth solution of an initial boundary value problem for the equations of a
homogeneous incompressible viscous MHD flow of zero resistivity without external forces in a bounded
domain (i.e., $\rho$ is a constant and $g=0$ in \eqref{0101}). Therefore, at present we can not
establish a nonlinear stability result for the more complicated problem \eqref{0103}--\eqref{0106}. On the other hand,
for the nonlinear instability problem, even if we have a local-in-time existence result for \eqref{0103}--\eqref{0106}, there still are
some difficulties induced by boundary in establishing the nonlinear instability from the linear instability
by a classical bootstrap argument \cite{GYSWIC,GYSWICNonlinea}. Due to the same reason, in a previous work \cite{JFJSWWWN}
we only considered a horizontally periodic domain with infinite height. We mention that for the Cauchy problem,
one does have certain existence results on global small smooth solutions \cite{LFZPGC}
and local large solutions \cite{HOCELETFCMDSFA}.
\end{rem}
\begin{rem}\label{rem:0206n} In this remark we give some extended results.
\begin{itemize}
 \item If $\Omega$ is a $C^{k+2}$-smooth bounded domain and $\bar{\rho}\in C^{k+1}(\bar{\Omega})$, then
     $(\tilde{u},\tilde{q})\in H^{k+2}\times  H^{k+1}$ by the classical regularity theory on the Stokes problem
     (see \cite[Theorem IV.6.1]{galdi2011introduction}).
 \item If $m=m_{\mm{C}}^i$, then we only have the following stability estimate of $u_t$, which can be observed from \eqref{engeryequality}.
\begin{equation}\label{criticase}
\|   {u}_t \|^2_{L^2}+ \mu \int_0^t \| {u}_\tau\|^2_{H^1}\mm{d}\tau
\leq  C\|(\varrho_0,\partial_i{u}_0,\mu\Delta {u}_0,\partial_i N_0)\|_{L^2}^2.
\end{equation}
\item If $\bar{\rho}$ is further in $C^2(\bar{\Omega})$, then we can deduce that $\partial_j(\bar{\rho}'\tilde{u}_3)\not\equiv 0$ for $j=1,2,3$,
and hence $\|\partial_j\varrho(t)\|_{L^2}\rightarrow \infty\mbox{ as }t\rightarrow \infty$. Moreover,
 based on \eqref{uestimeate2}, we can deduce from \eqref{0106} that the solution in the linear stability result satisfies
 $\| \partial_i\varrho (t) \|^2_{L^2}\leq C\|(\varrho_0,\partial_j\varrho_0,{u}_0,Q_0)\|_{L^2}^2$, whence,
$\lim_{t\rightarrow \infty}\|\partial_j\varrho(t)-\partial_j\varrho_\infty\|_{L^2}=0$, also see the derivation of \eqref{behaveioruofrho}.
\item Obviously, Theorem \ref{thm:0201} still holds for $i=2$ and $\bar{M}=(0,m,0)$.
            \end{itemize}
\end{rem}

Next, we sketch the main idea of the proof of Theorem \ref{thm:0201} to explain how to utilize the critical number,
and the detailed proof will be given in Section \ref{sec:03}. We start with the proof idea of the linear instability.
As in \cite{JFJSO2014}, we make the following ansatz of growing mode solutions to the linearized problem:
\begin{equation*}
{\varrho}( x,t)=\tilde{\varrho} ( x)e^{\Lambda t},\; u(x,t)=\tilde{u}( x)e^{\Lambda t},\;
q( x,t)=\tilde{q}( x)e^{\Lambda t},\; N( x,t)=\tilde{ N}( x)e^{\Lambda t}\;\;\mbox{ for some }\Lambda>0.
\end{equation*}
Substituting this ansatz into  \eqref{0106}, we get
\begin{equation}\label{timedependent}\left\{\begin{array}{l}
\Lambda\tilde{\varrho}+ \bar{\rho}'\tilde{u}_3=0, \\[1mm]
 \Lambda \bar{\rho}\tilde{u}  +\nabla (\tilde{q}+\lambda_0m\tilde{N}_i)=\mu\Delta\tilde{v}
 +\lambda_0 m\partial_i \tilde{N}  -\tilde{\varrho}g e_3,\quad \mm{div}\tilde{u}=0,\\[1mm]
\Lambda\tilde{ N}=m \partial_i \tilde{u},\quad \mathrm{div}\tilde{N}=0,\end{array}\right.  \end{equation}
and then eliminating $\tilde{\rho}$ and $\tilde{N}$ by using the first and third equations, we arrive at
the time-invariant boundary problem \eqref{0201} for $\tilde{u}$ and $\tilde{q}$. Then we apply a modified variational
method as in \cite{JFJSO2014} to construct a solutions of the eigenvalue problem \eqref{0201}.
This idea was used probably first by Guo and Tice to deal with an ODE problem arising in the construction of
unstable linear solutions \cite{GYTI2}. In view of the basic idea of the modified variational method, we modify \eqref{0201} as follows.
 \begin{equation}\label{0301}
\left\{   \begin{array}{l}
\alpha \bar{\rho}\tilde{u} =s\mu\Delta\tilde{u} +g\bar{\rho}'\tilde{u}_3e_3  +\nabla\tilde{p} +
   \lambda_0  m^2\partial_i^2 \tilde{u},  \quad  \mm{div}\tilde{u}=0 , \\[1mm]
\tilde{u}|_{\partial\Omega}= {0},    \end{array}  \right.     \end{equation}
where $s>0$ is a parameter, $\tilde{p}:= -( s \tilde{q}+\lambda_0m^2\partial_i
\tilde{u}_i) $ and $\alpha:=\alpha(s)$ depends on $s$.
 The system \eqref{0301} satisfies the identity: $\alpha(s)J(\tilde{u})= \mathcal{E}_\sigma(\tilde{u},s)$, where
$$\mathcal{E}_\sigma(\tilde{u},s):=E_\sigma(\tilde{u})- s \mu\int|\nabla\tilde{u}|^2\mm{d} x\;\mbox{ and }\;
E_\sigma(\tilde{u}):=\int g\bar{\rho}'\tilde{u}_3^2\mm{d} x -\lambda_0m^2\int|\partial_i\tilde{u}|^2\dx .$$
 Thus, using the variational method and the classical regularity theory on the Stokes problem, we find that $\tilde{u}\in H^2\cap \mathcal{A}_\sigma$, and
 $\alpha$ satisfies \eqref{0301} by maximizing
\begin{equation}\label{0206}
 \alpha(s)=\sup_{w\in\mathcal{A}_\sigma}\mathcal{E}_\sigma(w,s)\in \mathbb{R}.\end{equation}
 Moreover, in view of the definition of $\alpha(s)$, we can infer that $\alpha (s)\in C_{\mathrm{loc}}^{0,1}(0,\infty)$ is nonincreasing.
On the other hand, recalling the instability condition $|m|<m_{\mm{C}}^{ {i}}$, we conclude that
$E_\sigma( w_0)>0$ for some $w_0\in H^1_\sigma$, and thus there exists a finite interval $(0,\mathfrak{S})$ on which
$\alpha(s)>0$ and $\lim_{s\rightarrow \mathfrak{S}}\alpha(s)=0$. Employing a fixed-point argument, we immediately see
that there is a $\Lambda$ satisfying
\begin{equation} \label{growth}
 \Lambda=\sqrt{\alpha(\Lambda)}=\sqrt{\sup_{w\in\mathcal{A}_\sigma}\mathcal{E}_\sigma( w,\Lambda)} \in (0,\mathfrak{S}). \end{equation}
Consequently, we can construct a nontrivial solution $\tilde{u}\in H^2 $ to \eqref{0201}
  with a positive $\Lambda$ defined by \eqref{growth}, and therefore the linear instability follows.

For the linear stability, the key observation is that any solution to the linearized magnetic RT problem
\eqref{0104}--\eqref{0106} satisfies  \begin{equation}\label{engeryequality}
\int  \bar{\rho}| u_t|^2 \mathrm{d} x-E_\sigma(u(t))+
2\mu\int_0^t\int |\nabla u_\tau|^2\mm{d} x\mm{d}\tau=J_0,
\end{equation}
 where $J_0:=\int \left(\bar{\rho}| u_t|^2+\lambda_0m^2|\partial_i {u}|^2
-g\bar{\rho}'{u}_3^2\right)|_{t=0}\mathrm{d} x$. On the other hand, recalling the definition of the critical number, we get
 \begin{equation*}\label{varitaotainal}
g \int \bar{\rho}'{u}_3^2 \dx\leq { \lambda_0( m_{\mm{C}}^{i})^2\int |\partial_i{u}|^2 \dx},
\end{equation*}
whence,
\begin{equation}\label{varitaotainalnew}
-E_\sigma (u)\geq\lambda_0[m^2-( m_{\mm{C}}^{i})^2]\int|\partial_i {u}|^2 \mathrm{d} x. \end{equation}
Hence \eqref{uestimeate} follows from \eqref{varitaotainalnew},  \eqref{0106}$_1$, \eqref{0106}$_3$ and the condition
$|m|>m_{\mm{C}}^i$.

Furthermore, similarly to the derivation of \eqref{engeryequality}, we find that
\begin{equation}\label{newcompress}
\begin{aligned}&   \int \left[\bar{\rho}| u |^2-E_\sigma\left(\int_0^t{u}(\tau)\mm{d}\tau\right)\right]\mathrm{d} x +
2\mu\int_0^t\int |\nabla  u(\tau)|^2\mm{d} x\mm{d}\tau\\
&=\int  \bar{\rho}| u_0 |^2\mm{d}x+ 2\int_0^t\int Q_0 u\mm{d}x\mm{d}\tau\leq
\int  \bar{\rho}| u_0 |^2\mm{d}x+2\|Q_0\|_{L^2}\left\|\int_0^t u(\tau)\mm{d}\tau\right\|_{L^2},  \end{aligned}
\end{equation}
which, together with Cauchy-Schwarz's inequality, gives \eqref{uestimeate2} immediately. Finally, with these stability estimates in hand,
we can deduce \eqref{heighesimte}--\eqref{instabi} by standard energy estimates and asymptotic analysis.

\subsection{Compressible case}
Based the ideas in the proof of Theorem \ref{thm:0201}, we can
further establish the instability/stability of the linearized Parker problem \eqref{c0104}--\eqref{lincom}
by a more careful analysis for compressibility. Unfortunately, we can not give a critical number of the magnetic field for \eqref{c0104}--\eqref{lincom}
as for the incompressible problem \eqref{0104}--\eqref{0106},
since  the horizontal steady magnetic field is vertically stratified.
Now, we introduce another version of instability/stability criterion  for \eqref{c0104}--\eqref{lincom}.

First, substituting the following ansatz of growing mode solutions into \eqref{lincom},
\begin{equation*}
{\varrho}( x)=\tilde{\varrho}(x)e^{\Lambda t},\;\; u(x)=\tilde{u}( x)e^{\Lambda t},\;\;  N( x)=\tilde{N}( x)e^{\Lambda t},
\end{equation*}
we get
\begin{equation}\label{comlinear}\left\{\begin{array}{l}
\Lambda\tilde{\varrho}+\mm{div}( \bar{\rho}\tilde{u})=0, \\[1mm]
 \Lambda \bar{\rho}\tilde{u}　 +\nabla (p'(\bar{\rho})\tilde{\varrho}+
  \lambda_0m_{\mm{c}}\tilde{N}_1)=\mu\Delta\tilde{ {{v}}}
  +\mu_0\nabla\mm{div}{\tilde{u}}　 + \lambda_0\tilde{N}_3 \bar{ M}'_{\mm{c}}　 +\lambda_0
m_{\mm{c}}\partial_1\tilde{N}　 -\tilde{\varrho}g e_3,\\[1mm]
\Lambda\tilde{ N}=m_{\mm{c}}\partial_1 \tilde{u}
 -\tilde{u}_3 \bar{M}'_\mm{c}-\bar{M}_\mm{c}\mm{div}\tilde{u},\\[1mm]
\mathrm{div}\tilde{N}=0,\end{array}\right.  \end{equation}
and then eliminating $\tilde{\rho}$ and $\tilde{M}_{\mm{c}}$ by using (\ref{comlinear})$_1$ and (\ref{comlinear})$_2$, we arrive at a
time-independent boundary value problem problem for $\tilde{u}=(\tilde{u}_1,\tilde{u}_2,\tilde{u}_3)$:
                       \begin{equation}\label{com0201}
                            \left\{  \begin{array}{l}
\Lambda^2\bar{\rho}\tilde{u} =g\bar{\rho}'\tilde{u}_3e_3+\nabla [p'(\bar{\rho})\mm{div}
   ( \bar{\rho}\tilde{u})]+g\bar{\rho}\mm{div}\tilde{u} e_3+
   \lambda_0 m_{\mm{c}}(m_{\mm{c}}\partial_1^2 \tilde{u} -\bar{M}_{\mm{c}}\mm{div}\partial_1\tilde{u}) \\
   \qquad\qquad +\nabla [ \lambda_0m_{\mm{c}} (m_{\mm{c}}\partial_2 \tilde{u}_2+m_{\mm{c}}\partial_3 \tilde{u}_3+m'_{\mm{c}}
\tilde{u}_3) ]+\Lambda\mu\Delta\tilde{u}  +\Lambda\mu_0\nabla\mm{div}{\tilde{u}},\\
 \tilde{u}|_{\partial\Omega}= {0}. \end{array}\right.
\end{equation}

Multiplying \eqref{com0201} by $\tilde{u}$, integrating the resulting equations and using the condition \eqref{comsteady}, we infer that
\begin{align*}  \begin{aligned}
\Lambda^2\int\bar{\rho}|\tilde{u}|^2\mm{d} x = & \int[ g\bar{\rho}'\tilde{u}_3^2- p'(\bar{\rho})\bar{\rho}|\mm{div}\tilde{u}|^2  +
   \lambda_0 m_{\mm{c}}(m_{\mm{c}}\partial_1^2 \tilde{u}\cdot\tilde{u} -m_{\mm{c}}\tilde{u}_1 \mm{div}\partial_1\tilde{u}) \\
 &  \quad -  \lambda_0m_{\mm{c}} (m_{\mm{c}}\partial_2 \tilde{u}_2+m_{\mm{c}}\partial_3 \tilde{u}_3+m'_{\mm{c}}
\tilde{u}_3)  \mm{div}\tilde{u}\\
&\quad +g\bar{\rho}\tilde{u}_3\mm{div}\tilde{u}-p'(\bar{\rho})\bar{\rho}'\tilde{u}_3 \mm{div}\tilde{u}]\mm{d}x
    -\Lambda \int (\mu|\nabla\tilde{u}|^2+\mu_0|\mm{div}\tilde{u}|^2)\mm{d} x \\
=& \int\{ g\bar{\rho}'\tilde{u}_3^2- p'(\bar{\rho}) \bar{\rho}|\mm{div}\tilde{u}|^2 -\lambda_0 m^2_{\mm{c}} [|\partial_1\tilde{u}|^2 +
(\partial_2 \tilde{u}_2+ \partial_3 \tilde{u}_3)\mm{div}\tilde{u}- \partial_1\tilde{u}_1 \mm{div} \tilde{u}]\\
&\quad +(g\bar{\rho}-\lambda_0m_{\mm{c}} m'_{\mm{c}}-p'(\bar{\rho} )
    \bar{\rho}') \tilde{u}_3\mm{div}\tilde{u}\}\mm{d}x-\Lambda  \int (\mu|\nabla\tilde{u}|^2+\mu_0|\mm{div}\tilde{u}|^2)\mm{d} x\\
=& \int[ g\bar{\rho}'\tilde{u}_3^2- p'(\bar{\rho}) \bar{\rho}|\mm{div}\tilde{u}|^2
 -\lambda_0m^2_{\mm{c}}(|\partial_{1}\tilde{u}_2 |^2+|\partial_{1}\tilde{u}_3 |^2
+|\partial_{1}\tilde{u}_1-\mm{div}\tilde{u}|^2 ) \\
&\quad  +2 g\bar{\rho}\tilde{u}_3\mm{div}\tilde{u}]\mm{d}x-\Lambda  \int (\mu|\nabla\tilde{u}|^2+\mu_0|\mm{div}\tilde{u}|^2)\mm{d} x.
\end{aligned}
 \end{align*}

Defining
\begin{eqnarray*} &&
E_{\mathrm{c}}(\tilde{u}) :=\int g\bar{\rho}'\tilde{u}_3^2\mm{d} x +\int 2 g\bar{\rho}\mm{div}\tilde{u}\tilde{u}_3\dx
 -\int \Big[ p'(\bar{\rho})\bar{\rho}|\mathrm{div}\tilde{u}|^2 \\
 && \qquad\qquad + \lambda_0m^2_{\mm{c}}(|\partial_{1}\tilde{u}_2 |^2+|\partial_{1}\tilde{u}_3 |^2
+|\partial_{2}\tilde{u}_2+\partial_3\tilde{u}_3|^2  ) \Big] \dx ,  \\
&& V (\tilde{u})=: \int (\mu|\nabla\tilde{u}|^2+\mu_0|\mm{div}\tilde{u}|^2)\mm{d} x,
\end{eqnarray*}
we have the relation:
\begin{equation}\label{enclcomreo}
\Lambda^2J(\tilde{u})=\mathcal{E}_\mm{c}(\tilde{u},\Lambda):= E_\mm{c}(\tilde{u}) -\Lambda V (\tilde{u}).
\end{equation}
Recalling the proof of linear instability in Theorem \ref{thm:0201},
 the instability condition $|m|<m_{\mm{C}}^{ {i}}$ guarantees that
$E_\sigma(w)>0$ for some ${w}\in H^1_\sigma$. This gives the existence of a positive growth-rate $\Lambda$
by applying the modified variational method. Similarly, in view of \eqref{enclcomreo}, we shall impose the following condition:
\begin{equation}\label{instabilitycondiot}
\mbox{there exists a function }{w}\in H_0^1,\mbox{ such that }E_\mathrm{c}(w) >0,
\end{equation}
which is actually a necessary condition for the existence of a positive growth-rate $\Lambda$. Of course, we shall show
 that \eqref{instabilitycondiot} indeed is the instability condition for the linearized Parker problem in Subsection \ref{sect:0401}.

Now we turn to a sufficiently condition for the linear stability. In the proof of the linear stability in Theorem \ref{thm:0201},
the key step is to deduce the relation \eqref{varitaotainalnew}, from which and the condition $|m_{\mm{c}}|>m_{\mm{C}}^{ {i}}$
we obtain the desired stability estimates immediately. Define
$$ \mm{C_r}:=\sup_{w\in H_0^1} \frac{E_{\mm{c}}(w)}{\int
  \lambda_0 (|\partial_{1}w_2 |^2+|\partial_{1}w_3 |^2 +|\partial_{2}w_2+\partial_3{w}_3|^2 )  \mm{d}x},$$
where we have omitted to write the restriction that $w\in H_0^1$ should make the denominator and the square root operation sense.
Hence, if ${\mm{C_r}}<0$, we find that
\begin{equation}\label{comkestep}\begin{aligned} E_\mathrm{c}( {u})\leq\lambda_0\mm{C_r}
 \int (|\partial_{1}{u}_2|^2+|\partial_{1}{u}_3|^2 +|\partial_{2}{u}_2+\partial_3{u}_3|^2)\mm{d}x\;\;\mbox{ for any }u\in H_0^1.
\end{aligned}\end{equation}
Then we can deduce from \eqref{comkestep} the stability estimate on the velocity of solutions to
the linearized Parker problem, the detailed derivation will be presented in Subsection \ref{linearstabofPrar}.
In addition, it is easy to check that $\mm{C_r}>0$ is equivalent to \eqref{instabilitycondiot}; moreover $\mm{C_r}\neq +\infty$.

Now, we sum up the previous discussions and state the instability and stability results for the linearized Parker problem
\eqref{c0104}--\eqref{lincom}.
\begin{thm}\label{thm:0202}
 Assume that $\bar{\rho}$ and $m_{\mm{c}}$ satisfy \eqref{0102} and \eqref{com:01091}, and let $\bar{M}_{\mm{c}}=(m_{\mm{c}},0,0)$.
\begin{itemize}
\item[(1)] If $\mm{C_r}>0$,
then the equilibrium state $(\bar{\rho}, {0},\bar{M}_{\mm{c}})$ of the problem
\eqref{c0104}--\eqref{lincom}
 is unstable, that is, there exists an unstable solution
$$({\varrho}, {u},N):=e^{\Lambda t}(-\mm{div}(\bar{\rho}\tilde{u})/\Lambda,\tilde{ {u}},(m_{\mm{c}}\partial_1 \tilde{u}
 -\tilde{u}_3 \bar{M}_{\mm{c}}'-\bar{M}_{\mm{c}}\mm{div}\tilde{u})/\Lambda)$$
of \eqref{c0104}--\eqref{lincom}, where $\tilde{u}\in H^2\cap \mathcal{A}$ solves the boundary value problem \eqref{com0201}
  with a finite growth rate $\Lambda >0$ satisfying
\begin{equation}
\label{0111nn} \Lambda^2=\sup_{w\in\mathcal{A}}\mathcal{E}_{\mm{c}}(w,\Lambda)= {\mathcal{E}_{\mm{c}} (\tilde{u},\Lambda)}.
 \end{equation}
Moreover,
\begin{equation*}\label{0204ncomre}  \|(u_1,u_2)(t)\|_{L^2},\ \|u_3(t)\|_{L^2},\ \| (N_1,N_2)(t)\|_{L^2}\;\mbox{ and }\;\| N_3(t)\|_{L^2}\to
  \infty\mbox{ as }t\rightarrow \infty.
\end{equation*}
In addition, $\|\varrho(t)\|_{L^2}\to\infty$ as $t\to\infty$  provided $\bar{\rho}'\geq 0$.
  \item[(2)] If $\mm{C_r}<0$, and the initial data $\varrho_0,u_0,N_0$ satisfy
 $$u_0\in H^1_0\cap H^2,\;\varrho_0\in H^1,\; (N_0,\partial_1 N_0,\nabla N_1^0)\in L^2\times L^2\times L^2\;\mbox{ and }\; \mm{div}N_0=0,$$
then there exists a unique global solution $(\varrho,u, N)\in C^0(\mathbb{R}^+,H^1)\times C^0(\mathbb{R}^+ ,H^2)\times C^0(\mathbb{R}^+,L^2)$
of \eqref{c0104}--\eqref{lincom} satisfying the following stability estimates:
 \begin{align}  \label{comuestimeate}
&\|(\varrho, u, N)_t\|^2_{L^2}+\| ({u},\partial_1 u,\mm{div}u)(t)\|^2_{L^2}+ \mu\int_0^t\|u_\tau\|^2_{H^1}\mm{d}\tau\leq CI_0,\\
\label{uestimeate2c}
& \| u (t)\|^2_{L^2}+\left\|  \int_0^t (u, \partial_1 u,\mm{div}u)(\tau)\mm{d}\tau \right\|^2_{L^2}
+\mu \int_0^t \| {u}(\tau) \|^2_{H^1}\mm{d}\tau \leq  C(\|  ({u}_0,P_0)\|_{L^2}^2+ I_0),\\[1mm]
\label{heighesimtec}
 &\| (\varrho, N)(t) \|^2_{L^2} \leq  C(\| (\varrho_0,  {u}_0, N_0,P_0)\|_{L^2}^2+I_0), \\[1mm]
 \label{0324c}
  & \mu\|\nabla{ u}(t)\|_{L^2}^2\leq C(\| (\varrho_0,u_0,  N_0,P_0 )\|_{L^2}^2+I_0),
\end{align}
where $P_0=\nabla ( p'(\bar{\rho})\varrho_0+ \lambda_0
 m_{\mm{c}} N_1^0)- \lambda_0 N_3^0  \bar{M}_{\mm{c}}'- \lambda_0 m_{\mm{c}} \partial_1  N_0+\varrho_0  ge_3$, and  $I_0$ is defined in $\eqref{definforI0}$ and can be bounded from above by
\begin{equation}\label{bondCby}
C[\|\varrho_0\|_{H^1}+(\mu+\mu_0)\|\Delta u\|_{L^2}+\|(\partial_1u_1^0,\partial_2u_2^0,\partial_3u_3^0,N_3^0,\partial_1N_0,\nabla N_1^0)\|_{L^2} ].
\end{equation}
Furthermore, there is a couple $(\varrho_\infty, N_\infty)\in L^2$, such that
\begin{align}    & \label{instabic1}    \begin{aligned}
 &\int\big[\lambda_0 m_{\mm{c}}N_\infty\cdot \partial_1 \psi +g\varrho_\infty\psi_3 -\lambda_0 N_3^\infty m_{\mm{c}}\psi_1
-  (p'(\bar{\rho})\varrho_\infty +\lambda_0 m_{\mm{c}} N_1^\infty)\mm{div}\psi\big]\dx =0, \\
& \qquad \mbox{ for any }\psi=(\psi_1,\psi_2,\psi_3)\in H^1_0,   \end{aligned}   \\
\label{instabic}
&\|\varrho(t)-\varrho_\infty\|_{L^2},\  \| {u}(t)\|_{H^1},\ \|u_t\|_{L^2}\;\mbox{ and }\;\|N(t)-N_\infty\|_{L^2}\to 0\mbox{ as }t\to\infty,
\end{align}
where $N_j^\infty$ ($j=1,3$) denotes the $j$-th component of $N^\infty$.
\end{itemize}
\end{thm}
\begin{rem}
We can choose a sufficiently large $|m_{\mm{c}}|$, such that $\mm{C_r}<0$. In fact, we define
$$\bar{n}:=\sup_{w\in H_0^1} \sqrt{\frac{\int(g\bar{\rho}'w_3^2+2 g\bar{\rho}w_3\mm{div}w- p'(\bar{\rho})
     \bar{\rho}|\mm{div}w|^2 )\mm{d}x}{\int\lambda_0 (|\partial_{1}w_2 |^2+|\partial_{1}w_3 |^2
+|\partial_{2}w_2+\partial_3{w}_3|^2  )  \mm{d}x}}.$$
By Remark \ref{rem:0201}, we see that $\bar{n}\in (0,+\infty)$.
In view of \eqref{com:01091}, for given $p$, $\bar{\rho}$, $g$ and $\lambda_0$, we can choose $m_{\mm{c}}$ satisfying
$\inf_{x\in\Omega}|m_{\mm{c}}| >\bar{n}$. Then, it is easy to verify that such $m_{\mm{c}}$ satisfies $\mm{C_r}<0$.
  \end{rem}
\begin{rem} We give some extended results similar to those in Remark \ref{rem:0206n}:
\begin{itemize}
 \item
If $\Omega$ is a $C^{k+2}$-smooth bounded domain and $\bar{\rho}\in C^{k+1}(\bar{\Omega})$, then
$\tilde{u} \in H^{k+2}$ by the classical regularity theory on elliptic equations
(see \cite[Theorem 4.11]{AnintroudctuionGMML}).
\item Any linear solution of \eqref{c0104}--\eqref{lincom} satisfies the following estimate
$$\| \sqrt{\bar{\rho}}  {u}_t \|^2_{L^2}+ 2 \mu\int_0^t \|\nabla {u}_\tau\|^2_{L^2}\mm{d}\tau\leq I_0,$$
 provided $E_\mathrm{c}( {w})\leq 0$ for any $w\in H_0^1$ (i.e., $\mm{C_r}\leq 0$).
\end{itemize}
\end{rem}

\section{Proof of Theorem \ref{thm:0201}}\label{sec:03}
\subsection{Linear instability}\label{sec:0301}
In this Subsection we give the detailed proof of the linear instability in Theorem \ref{thm:0201}.
To begin with, we show that a maximizer of \eqref{0206} exists and that the corresponding Euler-Lagrange equations are
equivalent to \eqref{0301}.
\begin{pro}\label{pro:0201}
  Assume that the  density profile $\bar{\rho}$ satisfies the first two conditions in \eqref{0102}, then for any but fixed $s>0$, the following assertions are valid.
 \begin{enumerate}
    \item[(1)] $\mathcal{E}_\sigma(w,s)$ achieves its supremum on $\mathcal{A}_\sigma$.
   \item[(2)] Let $\tilde{ u} $ be a maximizer and $\alpha$ satisfy \eqref{0206},
then there exists a corresponding  pressure field $\tilde{p}$ associated with $\tilde{ u}$,
 such that the couple ($\tilde{ u} $, $\tilde{p} $) satisfies the boundary value problem \eqref{0301}. Moreover,
$(\tilde{ u} ,\tilde{p} ) \in H^2\times H^1$.
 \end{enumerate}
\end{pro}
\begin{pf}
(1)
Let $\{w_n\}_{n=1}^\infty\subset \mathcal{A}_\sigma$ be a maximizing sequence, then
$\mathcal{E}_\sigma(w_n,s)$ is  bounded.
Recalling that
$$ \mathcal\mathcal\mathcal{E}_\sigma(w_n,s)=\int[
g\bar{\rho}'(w_3^n)^2  -\lambda_0m^2|\partial_iw_n|^2 -s
 \mu|\nabla w_n|^2]\mm{d} x,$$
 we can easily see that $w_n$ is bounded in $H^1_\sigma$.
 Here $w_3^n$ denotes the third component of $w_n$.
So, there exists a
$\tilde{w}\in  \mathcal{A}_\sigma$ and a subsequence (still denoted by $ w_n$ for simplicity), such that
$w_n\rightarrow \tilde{ w} $ weakly in
$H^1_\sigma$ and strongly in $L^2$. Moreover, by the lower semi-continuity, one has
\begin{equation*}
\begin{aligned}
\sup_{ {w}\in \mathcal{A}_\sigma}\mathcal\mathcal\mathcal{E}_\sigma( { w},s)= &\limsup_{n\rightarrow
\infty}\mathcal\mathcal\mathcal{E}_\sigma(w_n,s)
\\
= &\lim_{n\rightarrow \infty}\int \bar{\rho}' ({w}_{3}^n)^2\mm{d} {x}-
\liminf_{n\rightarrow \infty} \int ( \lambda_0m^2|\partial_i w_n|^2+s\mu|\nabla w_n|^2)\mm{d} {x}\\
\leq & \mathcal\mathcal\mathcal{E}_\sigma(\tilde{w} ,s)\leq \sup_{ { w}\in\mathcal{A}_\sigma}\mathcal\mathcal\mathcal{E}_\sigma( { w},s),
\end{aligned}\end{equation*}
which shows that $\mathcal\mathcal\mathcal{E}_\sigma(w,s)$ achieves its supremum on $\mathcal{A}_\sigma$.

(2) To show the second assertion, we notice that since $\mathcal{E}_\sigma(w,s)$ and $J(w)$
 are homogeneous of degree $2$, \eqref{0206} is equivalent to
  \begin{equation}\label{inc0227}
\alpha=\sup_{w\in {H}^1_{\sigma}}\frac{\mathcal\mathcal\mathcal{E}_\sigma(w,s)}{J(w)}.
\end{equation}
For any $\tau\in \mathbb{R}$ and $ \varphi\in {H}^1_\sigma$, we take
$\tilde{v}(\tau):=\tilde{ {u}} +\tau \varphi$, where  $\tilde{ u} $ is
a maximizer. Then \eqref{inc0227} implies
  \begin{equation*}\mathcal\mathcal\mathcal{E}_\sigma(\tilde{ v}(\tau),s)-\lambda^2J(\tilde{v}(\tau))\leq 0.
\end{equation*}
If we set
$I(\tau)=\mathcal\mathcal\mathcal{E}_\sigma(\tilde{v}(\tau),s)-\Lambda^2J(\tilde{v}(\tau))$,
then we see that $I(\tau)\in C^1(\mathbb{R})$, $I(\tau)\leq 0$ for all $\tau\in \mathbb{R}$ and $I(0)=0$.
This implies $$\frac{\mm{d} I(\tau)}{\mm{d}\tau}\bigg|_{\tau=0}=0.$$ Hence, a direct computation leads to
  \begin{equation*}\begin{aligned}
\alpha \int \bar{\rho}\tilde{u}\cdot \varphi\mm{d}x
=&-s\mu\int\nabla \tilde{ {{u}}}:\nabla \varphi\mm{d}x
 -   \lambda_0 m^2\int
\partial_i\tilde{u} \cdot\partial_i \varphi \mm{d} {x}+\int g\bar{\rho}'\tilde{u}_{3}e_3 \cdot \varphi\mm{d}x ,\end{aligned}  \end{equation*}
which implies that $\tilde{ {u}}\in H^1_\sigma $ is a weak solution to \eqref{0301}.
Thus, by virtue of scaling the variable $x_i$, it follows from the classical regularity theory on the Stokes problem
 that there are constants $c_1$ dependent of the domain $\Omega$, $s$, $\mu$, $\lambda_0$ and $m$, and $c_2$ dependent
 of $c_1$, $g$, $\alpha$ and $\bar{\rho}$, such that
  \begin{equation*}
  \|\tilde{ {u}}\|_{H^{2}} +\|\nabla \tilde{p}\|_{L^{2}}\leq c_1\|
(g\bar{\rho}'\tilde{{u}}_{3} {e}_3-\alpha\bar{\rho}\tilde{ {u}}) \|_{L^2}\leq c_2,
\end{equation*}
  where $\tilde{p}  \in H^{1}$ is the corresponding pressure field associated to $\tilde{u}$. This completes the proof.
\hfill $\Box$
\end{pf}

Next, we want to show that there is a fixed point $s=\Lambda>0$ such that $\sqrt{\alpha(\Lambda)}=\Lambda$. To this end,
we first give some properties of $\alpha(s)$ as a function of $s> 0$.
\begin{pro}\label{pro:0202}
Under the assumptions of Proposition \ref{pro:0201}, the
function $\alpha(s)$ defined on $(0,\infty)$ enjoys the following properties:
\begin{enumerate}[\quad \ (1)]
 \item $\alpha(s)\in C_{\mathrm{loc}}^{0,1}(0,\infty)$ is nonincreasing.
   \item
If $|m|<m_{\mathrm{C}}^{i}$, then there are constants $c_3$, $c_4>0$ which depend on $g$, $\bar{\rho}$, $\mu$, $\lambda_0$ and $m$, such that
  \begin{equation}\label{n0302}\alpha(s)\geq c_3-sc_4
  .\end{equation}
  \end{enumerate}
\end{pro}
\begin{pf}
(1) Let $\{w_n^{s_j}\}_{n=1}^\infty\subset\mathcal{A}_\sigma$ be a maximizing sequence
of $\sup_{w\in\mathcal{A}_\sigma}\mathcal{E}_\sigma(w,s_j)$ for $j=1$ and $2$. Then
\begin{equation*}
\alpha(s_1)\geq
\limsup_{n\rightarrow\infty}\mathcal\mathcal\mathcal{E}_\sigma(w^{s_2}_n,s_1) \geq
\liminf_{n\rightarrow\infty}\mathcal\mathcal\mathcal{E}_\sigma(w^{s_2}_n,s_2)=\alpha(s_2)\;
\mbox{ for any }0<s_1<s_2<\infty.
\end{equation*}
Hence $\alpha(s)$ is nonincreasing on $(0,\infty)$. Next we use this fact to show the continuity of $\alpha(s)$.

Let $I:=[a,b]\subset (0,\infty)$ be a bounded interval. In view of
  the monotonicity of $\alpha(s)$, we know that
\begin{equation} \label{n0303}
|\alpha(s)|\leq \max\left\{|\alpha(a)|,{g}  \left\|\bar{\rho}'/{\bar{\rho}}
\right\|_{L^\infty}\right\}<\infty\quad\mbox{ for any }s\in I.
\end{equation}
On the other hand, for any $s\in I$, there exists a maximizing sequence
$\{w_n^{s}\}_{n=1}^\infty\subset\mathcal{A}_\sigma$ of $\sup_{w\in
\mathcal{A}_\sigma}\mathcal\mathcal\mathcal{E}_\sigma(w,s)$, such that
\begin{equation}\label{n0304}\begin{aligned}|\alpha(s)
-\mathcal\mathcal\mathcal{E}_\sigma(w^{s}_n,s)|<1
\end{aligned}.\end{equation}
Making use of 
 \eqref{n0303} and \eqref{n0304}, we infer that
\begin{equation*}\begin{aligned} 0\leq & \mu \int|\nabla w|^2\mm{d} {x}+\frac{\lambda_0m^2}{s}\int|\partial_iw|^2\dx \\
=&\frac{g }{s}\int \bar{\rho}'|w^{s}_{n3}|^2\mathrm{d} {x} -\frac{\mathcal\mathcal\mathcal{E}_\sigma(w^s_n,s)}{s} \\
\leq & \frac{1+\max\{|\alpha(a)|,{g}\left\|{\bar{\rho}'}/{\bar{\rho}}
\right\|_{L^\infty}\}}{a}+\frac{g}{a}\left\|\frac{\bar{\rho}'}{\bar{\rho}}
\right\|_{L^\infty }:=K,
\end{aligned}\end{equation*}
where $w^{s}_{n3}$ denotes the third component of $w^{s}_{n}$.
Thus, for $s_j\in I$ ($j=1,2$), we further find that
\begin{equation}\begin{aligned}\label{0235}\alpha(s_1)= \limsup_{n\rightarrow
\infty}\mathcal\mathcal\mathcal{E}_\sigma(w^{s_1}_n,s_1)\leq & \limsup_{n\rightarrow
\infty}\mathcal\mathcal\mathcal{E}_\sigma(w^{s_1}_n,s_2)+\mu
|s_1-s_2|\limsup_{n\rightarrow
\infty}\int |\nabla w^{s_1}_n|^2\mm{d} {x}\\
\leq & \alpha(s_2)+K|s_1-s_2|.
\end{aligned}\end{equation}
Reversing the role of the indices $1$ and $2$ in the derivation of the inequality
\eqref{0235}, we obtain the same boundedness with the indices switched. Therefore, we deduce that
\begin{equation*}\begin{aligned}|\alpha(s_1)-\alpha(s_2)|\leq K|s_1-s_2|,
\end{aligned}\end{equation*}
which yields $\alpha(s)\in C_{\mathrm{loc}}^{0,1}(0,\infty)$.

(2) We turn to the proof \eqref{n0302}. Noting that $m<m_{\mm{C}}^{i}$, we can deduce
from the definition of $m_{\mm{C}}^{i}$ that there is a $ v\in H^1_\sigma$, such that
\begin{align*}
 \int g\bar{\rho}' v_3^2\mm{d} x-\lambda_0m^2\int|\partial_iv|^2\dx>0.   \end{align*}
Thus, one has
\begin{equation*}\begin{aligned}
\alpha(s)=& \sup_{w\in\mathcal{A}_\sigma}\mathcal\mathcal\mathcal{E}_\sigma(w,s)
=\sup_{w\in H_\sigma^1}\frac{\mathcal\mathcal\mathcal{E}_\sigma(w,s)}{J(w)} \\
& \geq \frac{\mathcal\mathcal\mathcal{E}_\sigma(v,s)}{J(v)}=
\frac{\int g\bar{\rho}'{v}_3^2\mm{d} x-\lambda_0m^2\int|\partial_i{ v}|^2\dx}{\int\bar{\rho}{v}^2\mm{d} {x}}
-s\frac{ \mu \int|\nabla v|^2\mm{d} {x}}{\int\bar{\rho}{v}^2\mm{d} {x}}:= c_3-sc_4
\end{aligned}\end{equation*} for two positive constants $c_3:
 =c_3(g,\bar{\rho},m)$ and $c_4:=c_4( \mu,\bar{\rho})$.
This completes the proof of Proposition \ref{pro:0202}.
 \hfill $\Box$
\end{pf}

Next we show that there exists a pair of functions $(\tilde{{u}},\tilde{q})$ satisfying
\eqref{0201} with a growth rate $\Lambda>0$.  Let
\begin{equation*}\label{0240}\mathfrak{S} :=\sup\{s~|~\alpha(\tau)>0\mbox{ for any }\tau\in (0,s)\}.
\end{equation*}
By virtue of Proposition \ref{pro:0202}, $\mathfrak{S}>0$; moreover, $\alpha(s)>0$ for any $s<\mathfrak{S}$ (in addition,
we can further show that $\alpha(s)$ strictly decreases on $(0,\mathfrak{S})$).
Since $\alpha(s)=\sup_{w\in\mathcal{A}_\sigma}\mathcal{E}_\sigma(w,s)<\infty$,
using the monotonicity of $\alpha(s)$, we see that
 \begin{equation}\label{zero}
 \lim_{s\rightarrow 0}\alpha(s)\mbox{ exists and the limit is a positve constant.}
 \end{equation}
 On the other hand, by virtue of Poincar\'{e}'s inequality, there is a constant $c_5$ dependent of
$g$, $\bar{\rho}$ and $\Omega$, such that
 $$g\int \bar{\rho}'w_3^2\mm{d} {x} \leq c_5\int|\nabla w|^2\mm{d} {x}\quad\mbox{ for any } w\in\mathcal{A}.$$
 Thus, if $s>c_5/\mu$, then
$$ g\int \bar{\rho}'w_3^2\mm{d} {x}-s\mu\int|\nabla w|^2\mm{d} {x}<0\quad\mbox{ for any } w\in\mathcal{A},$$
which implies that
 $$\alpha(s)\leq 0\quad \mbox{ for any } s>c_5/\mu. $$
Hence $\mathfrak{S}<\infty$. Moreover,
\begin{equation}\label{zerolin}
\lim_{s\rightarrow \mathfrak{S}}\alpha(s)=0.
\end{equation}

Now, exploiting \eqref{zero}, \eqref{zerolin} and the continuity of $\alpha(s)$ on $(0,\mathfrak{S})$,
we find by a fixed-point argument on $(0,\mathfrak{S})$ that there is a unique $\Lambda\in(0,\mathfrak{S})$ satisfying \eqref{growth}.
Thus, by virtue of Proposition \ref{pro:0201}, there is a solution
$(\tilde{ {u}},\tilde{q})\in H^2\times H^1$
to the problem  \eqref{0201} with $\Lambda$ constructed in
\eqref{growth}, where $\tilde{q}:=-(\tilde{p}+ \lambda_0m^2\partial_i
\tilde{u}_i)/ \Lambda$. We conclude the following proposition, which, together with Remark \ref{remark0204}, yields
 the linear instability in Theorem \ref{thm:0201}.
\begin{pro}\label{pro:nnn0203}
 Assume that the density profile $\bar{\rho}$ satisfies
\eqref{0102} and $m<m_{\mm{C}}^{\mm{i}}$. Then there exists a pair of functions
$(\tilde{ {u}},\tilde{q})\in (H^{2}\cap \mathcal{A}_\sigma)\times H^{1}$ which solves the boundary value problem
\eqref{0201}
with a finite growth rate $\Lambda>0$ satisfying \eqref{Lambdard}.
 In particular, $\tilde{\varrho}:=-\bar{\rho}'\tilde{{u}}_3/\Lambda\in H^1$ and $\tilde{N}:=m\partial_i \tilde{u}/\Lambda\in L^2$.
 Moreover, $(\tilde{\varrho},\tilde{ {u}},\tilde{N},\tilde{q})$ satisfies
\eqref{timedependent}.
\end{pro}
\begin{rem}
It is easy to see that Proposition \ref{pro:nnn0203} still holds for $\bar{\rho}$ being a function of three variables $(x_1,x_2,x_3)$.
However, if
$$\nabla \bar{p}=-\bar{\rho} g e_3,$$
then $\bar{\rho}$ has to be a function of the single variable $x_3$. This is the reason why we only consider that $\bar{\rho}$
is a single variable of $x_3$ in Theorem \ref{thm:0201}.
\end{rem}

\subsection{Linear stability}\label{stability}
Before proving the linear stability in Theorem \ref{thm:0201}, we shall establish the local well-posedness of the linearized magnetic
RT problem \eqref{0104}--\eqref{0106}, which can be shown by an iterative method.
Next, we briefly describe how to show it for the reader's convenience.

Let $T^*\in (0,1)$, $I_{T^*}:=(0,T^*)$, $K>0$ and
$$U_K=\{u\in C^0(\bar{I}_{T^*}, H^2 )\cap C^0(\bar{I}_{T^*}, H^1_\sigma )~|~\|u\|_{L^\infty(I_{T^*},H^2)} \leq K\},$$
where ${T^*}$ and $K$ will be fixed later. Given $v\in U_K$, we consider the following linear problem:
\begin{equation}\label{newsytem}
\left\{\begin{array}{ll}
   \bar{\rho} u_t +\nabla  {p}=\mu\Delta{ u}+f,\quad\mathrm{div} {u}=0,\\[1mm]
u|_{\partial\Omega}=0,\quad u|_{t=0}=u_0,
\end{array}\right.\end{equation}
where
\begin{align}\label{realtiopfNrgho}
f= \lambda_0 m\partial_{i}N-\varrho ge_3,\;\;\varrho=-\int_0^t\bar{\rho}'{v}_3(\tau)\mm{d}\tau+\varrho_0,\;\; N=m\int_0^t \partial_iv(\tau)\mm{d}\tau+N_0.
\end{align}
Obviously, $\mm{div} N=0$, since $\mm{div}N_0=0$. Using \eqref{realtiopfNrgho}, we have $(f,f_t)\in C^0(\bar{I}_{T^*}, L^2)$,
and thus the problem \eqref{newsytem} possesses a unique solution $u\in L^\infty( {I}_{T^*}, H^2)$ with a unique associated pressure
${p}\in L^\infty( {I}_{T^*}, L^2)$ satisfying $\int {p}\dx=0$. Moreover, $u$ and $p$ enjoy the following estimates
(referring to \cite[Lemma 5]{CYKHOM1ufda} or \cite[Lemma 5.2]{JFJSWWWN})
\begin{equation}\label{esimta1}\begin{aligned}
&\|u_t \|_{C^0(\bar{I}_{T}^*,L^2)}^2 + \|u\|_{L^\infty({I}_{{T^*}}, H^2)}^2+\|u_t\|_{L^2(I_{T^*},H^1 )}^2+\| {p}\|_{L^\infty({I}_{{T}^*},H^1)}^2\\
&\leq C_{\mu}[\|u_0\|_{H^2}^2+\|f_0\|_{L^2}^2+(1+T^*)\|f\|_{L^\infty(I_{T^*},L^2)}^2 +T^*\|f_t\|_{L^\infty(I_{T^*},L^2)}^2]\\
&\leq \tilde{C}(\|\varrho_0\|_{L^2}^2+ \|u_0\|_{H^2}^2+\|\partial_i N_0\|_{L^2}^2+K^2T^*)\quad\;\mbox{ for any }t\in I_{T^*},
\end{aligned}
\end{equation}
where the constant $\tilde{C}$ only depends on $\mu$, $\bar{\rho}$, $\lambda_0$, $g$, $m$ and the domain $\Omega$.
Using the classical regularity theory on the Stokes problem, we have
$$\|u(t_2)- u(t_1)\|_{H^2}+\|p(t_2)-  p(t_1)\|_{H^1}\leq C_\mu(\|u_t(t_2)-u_t(t_1)\|_{L^2}+\|f(t_2)-f(t_1)\|_{L^2}),$$
 which implies that  $u\in C^0(\bar{I}_{{T}^*}, H^2)$ and $ {p}\in C^0(\bar{I}_{{T}^*}, H^1)$.
Now, taking $K\geq\tilde{C}(\|\varrho_0\|_{L^2}^2+ \|u_0\|_{H^2}^2+\|\partial_i N_0\|_{L^2}^2+1)$ and $T^*\leq 1/K^2$, we arrive at $u\in U_K$.

Considering the above results, we can construct  a function sequence $\{u\}_{n=1}^\infty$ satisfying
\begin{equation*}\label{ajppnewsytem}
\left\{\begin{array}{ll}
   \bar{\rho} \partial_tu_{n+1} +\nabla  {p}_{n+1}=\mu\Delta{ u}_{n+1} +  \lambda_0 m\partial_{i}N_n-\varrho_n ge_3,\quad\mathrm{div} {u}_{n+1}=0,\\[1mm]
u_{n+1}|_{\partial\Omega}=0,\quad u_{n+1}|_{t=0}=u_0
\end{array}\right.\end{equation*}
and
\begin{equation*}\label{aud1}\|\partial_tu_n\|_{C^0(\bar{I}_{{T}^*},
L^2)}^2+
\|u_n\|_{C^0(\bar{I}_{{T}^*}, H^2)}^2+\|\partial_t u_n\|_{L^2(I_{T^*},H^1 )}^2+\|  {p}_n\|_{C^0(\bar{I}_{{T}^*},H^1)}\leq K,
\end{equation*}
where
$$\varrho_n=-\int_0^t\bar{\rho}'{v}_{3}^n(\tau)\mm{d}\tau+\varrho_0,\quad N_n=m\int_0^t \partial_iv_n(\tau)\mm{d}\tau+N_0. $$

Let $\bar{u}_{n+1}=u_{n+1}-u_n$, $\bar{\varrho}_{n}=\varrho_{n}-\varrho_{n-1}$, $\bar{N}_{n}=N_{n}-N_{n-1}$ and $\bar{p}_{n+1}= {p}_{n+1}- {p}_n$ for $n\geq 2$, then we have
\begin{equation*}\label{ajppnewsytemdife}
\left\{\begin{array}{ll}
   \bar{\rho} \partial_t\bar{u}_{n+1} +\nabla \bar{{p}}_{n+1}=\mu\Delta\bar{ u}_{n+1}+
  \lambda_0 m\partial_{i}\bar{N}_n-\bar{\varrho}_n ge_3,\quad  \mathrm{div} \bar{u}_{n+1}=0,\\[1mm]
\bar{u}_{n+1}|_{\partial\Omega}=0,\quad \bar{u}_{n+1}|_{t=0}=0\end{array}\right.\end{equation*}
In view of the estimate \eqref{esimta1}, we get
$$\begin{aligned}&\|\partial_t\bar{u}_{n+1}\|_{L^\infty({I}_{{T}^*},
L^2)}^2+\|\bar{u}_{n+1}\|_{L^{\infty}( {I}_{T^*},H^2)}^2
\\
&+\|\partial_t \bar{u}_{n+1}\|_{L^2(I_{T^*},H^1 )}^2+\|  \bar{p}_{n+1}\|_{L^\infty({I}_{{T}^*},H^1)}\leq  \tilde{C}T^{*}\|\bar{u}_{n}\|_{L^{\infty}( {I}_{T^*},H^2)}^2.
\end{aligned}$$
Hence, we see that $\{u_n\}_{n=1}^{\infty}$, $\{\partial_t u_n\}_{n=1}^{\infty}$ and $\{ {p}_n\}_{n=1}^{\infty}$ are a Cauchy sequence in
$$L^\infty( {I}_{T^*},H^2),\ L^2({I}_{T^*},H^1)\cap L^\infty({I}_{T^*},L^2)\;\mbox{ and }\; L^\infty( {I}_{T^*},H^1)$$ for
sufficiently small $T$, respectively. Thus we obtain the limit functions $u$, $p$. It is easy to verify that the limit functions $u$, $p$
are a unique solution to the following problem
\begin{equation}\label{newsytem12}
\left\{\begin{array}{ll}
   \bar{\rho} u_t +\nabla  {p}=\mu\Delta{ u}+ \lambda_0 m\partial_{i}N-\varrho ge_3,\quad  \mathrm{div} {u}=0,\\[1mm]
\varrho=-\int_0^t\bar{\rho}'{u}_3(\tau)\mm{d}\tau+\varrho_0,\quad N=m\int_0^t \partial_iu(\tau)\mm{d}\tau+N_0,\\
u|_{\partial\Omega}=0,\quad u|_{t=0}=u_0.
\end{array}\right.\end{equation}
Furthermore,
\begin{equation*}u\in C^0(\bar{I}_{T^*},
H^2),\   u_t\in C^0(\bar{I}_{T^*},L^2),\ \int p\dx=0\;\mbox{ and }\; {p}\in C^0(\bar{I}_{T^*},H^1).
\end{equation*}
 Obviously, $(\varrho, u, N)$ constructed above also uniquely solves the linearized problem \eqref{0104}--\eqref{0106} with an associated pressure
$q=p -\lambda_0 m N_i$. Moreover, $(\varrho,N,\partial_i N)\in C^0(\bar{I}_{T^*},L^2)$ due to $(\varrho_0,N_0,\partial_i N_0)\in L^2$.

To get a global solution  in Theorem \ref{thm:0201}, it suffices to deduce
the global estimates for $\|\varrho(t)\|_{L^2}$, $\|u(t)\|_{H^2}$ and $\|\partial_iN(t)\|_{L^2}$.
To begin with, we derive the energy equality \eqref{engeryequality}.
In view of the regularity of $(\varrho,u, N)$, we can deduce from \eqref{newsytem12}$_1$ that for a.e. $t>0$,
\begin{equation*}\label{n0310}
\begin{aligned}
\frac{1}{2}\frac{\mm{d}}{\mm{d}t}\int\bar{\rho}| u_t|^2
\mathrm{d} x= <\bar{\rho}{u}_{tt},{u}_t >  =\int(\lambda_0m^2 \partial_i^2 u\cdot u_t+ g\bar{\rho}' u_3
 \partial_t u_3)\mm{d}x- \mu\int |\nabla u_t|^2 \mm{d} x,
\end{aligned}\end{equation*}
where  $<\cdot,\cdot>$ denotes the dual product between the spaces $H^{-1}_\sigma $ and $H_\sigma^1 $, and $H^{-1}_\sigma $ represents the dual space of
$H_\sigma^1$,
please refer to   \cite[Remark 6]{CYKHOM1ufda}. On the other hand,
\begin{equation*} \begin{aligned}
\frac{1}{2}\frac{\mm{d}}{\mm{d}t}
\int\left(\lambda_0 m^2|\partial_i {u}|^2
-g\bar{\rho}'{u}_3^2\right)\mathrm{d} x &=\int\left( \lambda_0m^2  \partial_i {u} \cdot \partial_i {u}_t
-g\bar{\rho}'{u}_3\partial_t u_3\right)\mathrm{d} x\\
&= -\int\left(g\bar{\rho}'{u}_3\partial_t u_3+\lambda_0 m^2 \partial_i^2 u\cdot {u}_t
\right)\mathrm{d} x,
\end{aligned}\end{equation*}
Putting the previous two equalities together, we conclude
\begin{equation*}
\frac{\mm{d}}{\mm{d}t}\int\left(\bar{\rho}| u_t|^2+\lambda_0 m^2|\partial_i {u}|^2
-g\bar{\rho}'{u}_3^2\right)\mathrm{d} x + 2\mu \int |\nabla u_t|^2\mm{d} x =0,
\end{equation*}
which yields \eqref{engeryequality}.

Then, using the inequality \eqref{varitaotainalnew}, we further infer from \eqref{engeryequality} that
 \begin{equation*}
\int \left\{\bar{\rho}| u_t|^2+\lambda_0[m^2-( m_{\mm{C}}^{i})^2] |\partial_i  {u}|^2
 \right\}\mathrm{d} x + 2\mu\int_0^t\int |\nabla u_\tau|^2\mm{d} x\mm{d}\tau\leq J_0.
\end{equation*}
Recalling \eqref{poincareinequality} and Poinc\'are's inequality, we obtain
  \begin{equation}\label{compreuestimeate}
\|   {u}_t \|^2_{L^2}+\| ({u},\partial_i u) \|^2_{L^2}+\mu \int_0^t \| {u}_\tau\|^2_{H^1}\mm{d}\tau
\leq  CJ_0.
\end{equation}
Applying (\ref{compreuestimeate}) to the first and third equations in \eqref{0106}, we find that
\begin{equation}\label{uestimeate1}
\| (\varrho, {u}, N)_t \|^2_{L^2}+\mu\| ({u},\partial_i u) \|^2_{L^2} + \mu\int_0^t \| {u}_\tau\|^2_{H^1}\mm{d}\tau \leq  CJ_0.
\end{equation}
On the other hand, we have
\begin{equation}\label{intiestimate}
J_0\leq \|(\varrho_0, \partial_i {u}_0,\mu\Delta {u}_0,\partial_i N_0)\|^2_{L^2}.
\end{equation}
  In fact, using the second condition in \eqref{0102} and Cauchy-Schwarz's inequality, we obtain from \eqref{0106}$_2$ that
   \begin{equation*}  \begin{aligned}
    \int \bar{\rho}| u_\tau|^2(\tau) \mm{d} x=&
\int \left(\mu \Delta  u+\lambda_0 m\partial_i N -\varrho g {e}_3\right)\cdot {u}_{\tau}(\tau)\mm{d} x\\
\leq  & C(\|\varrho(\tau)\|_{L^2}^2 +\|\mu\Delta  {u}(\tau)\|_{L^2}^2+\|\partial_i N(\tau)\|_{L^2}^2)
+\frac{\inf{\bar{\rho}}}{2} \| u_\tau (\tau)\|^2_{L^2},
\end{aligned} \end{equation*}
which implies
 \begin{equation*} \begin{aligned}
      \|u_\tau(\tau)\|^2_{L^2} \leq C(\|\varrho(\tau)\|_{L^2}^2+\|\mu\Delta  {u}(\tau)\|_{L^2}^2
      +\|\partial_i N(\tau)\|_{L^2}^2) \quad \mbox{ for any }\tau>0.  \end{aligned} \end{equation*}
Let $\tau\rightarrow 0$, then
 \begin{equation*} \|u_t|_{t=0}\|^2_{L^2} \leq C\|(\varrho_0,\mu\Delta {u}_0,\partial_i N_0)\|_{L^2}^2,
 \end{equation*}
 and one gets \eqref{intiestimate}. Putting \eqref{uestimeate1} and \eqref{intiestimate} together, we obtain the stability estimate \eqref{uestimeate}.

Noting that $\partial_iN=m\int_0^t\partial_i^2u(\tau)\mm{d}\tau+\partial_iN_0$, thus, we use the classical regularity theory on the Stokes problem to
infer that
$$ \begin{aligned}
 \|u(t)\|_{H^2}+\| q(t)\|_{H^1}\leq &C_{\mu}(\|u_t (t)\|_{L^2} +\|\lambda_0 m\partial_{i}N-\varrho ge_3\|_{L^2})\\
\leq &C_{\mu}  \left(\|(\varrho_0,\partial_iu_0,\Delta {u}_0,\partial_i N_0)\|_{L^2}+\int_0^t\| u(\tau)\|_{H^2}\mm{d}\tau\right) ,
\end{aligned}   $$
which, together with Grownwall's inequality, yields
\begin{equation}\label{gloableestima}
\begin{aligned}
\|u(t)\|_{H^2}\leq   C_{\mu}\|(\varrho_0,\partial_i u_0,\Delta {u}_0,\partial_i N_0)\|_{L^2}\left(1+C_{\mu}t e^{C_{\mu}t}\right).
\end{aligned}
\end{equation}
With the global estimate \eqref{gloableestima} and
\begin{equation*}  \begin{aligned}
 \|(\varrho, \partial_i N)(t)\|\leq  & C  \left(\|(\varrho_0,\partial_i N_0 )\|_{L^2}+\int_0^t\| u(\tau)\|_{H^2}\mm{d}\tau\right)
\end{aligned}   \end{equation*}
in hand, we immediately get the global
solution $(\varrho, u,N)$ with an associated pressure $q$ by a continuity argument based on the local well-posedness result.
Moreover, the global solution satisfies the stability estimate \eqref{uestimeate}.

Now, we proceed to deriving the estimates \eqref{uestimeate2}--\eqref{0324}. 
 Firstly, \eqref{0106}$_2$ yields
\begin{equation*}\label{n0323n} \bar{\rho}  u_{t}+ \nabla  (q+ \lambda_0 m N_i)=
 \mu \Delta u+  \lambda_0m^2 \partial_i^2 \int_0^t u(\tau)\mm{d}\tau
 +\bar{\rho}' \int_0^t u_3(\tau) \mm{d}\tau g e_3+Q_0.
\end{equation*}
Consequently,
\eqref{newcompress} follows. Noting that $\int_0^t u(\tau)\mm{d}\tau\big|_{\partial\Omega}=0$, similarly to the derivation of \eqref{compreuestimeate},
we can obtain \eqref{uestimeate2}. Hence, exploiting \eqref{0106}$_1$, we have
\begin{equation}\begin{aligned}\label{n0324}
\|\varrho(t)\|_{L^2}\leq & \|\varrho_0\|_{L^2}+ \left\|\int_0^t\varrho_\tau \mm{d}\tau\right\|_{L^2} \\
\leq &\|\varrho_0\|_{L^2} +\|\bar{\rho}'\|_{L^\infty}\left\|\int_0^tu_3(\tau) \mm{d}\tau\right\|_{L^2}\leq C\|(\varrho_0,u_0,Q_0)\|_{L^2}.
\end{aligned}\end{equation}
 Similarly, using \eqref{0106}$_3$, one obtains
\begin{equation}\label{magneticstabtiliy}\begin{aligned}
\| N(t)\|_{L^2}\leq&
\| N_0\|_{L^2}+\left\|\int_0^t N_\tau\mm{d}\tau\right\|_{L^2}\\
\leq &\| N_0\|_{L^2}+|m|\left\|\int_0^t \partial_i{u}(\tau)\mm{d}\tau\right\|_{L^2}
\leq   C \|(u_0, N_0,Q_0)\|_{L^2}.
\end{aligned}\end{equation}
Therefore, the estimate \eqref{heighesimte} follows from \eqref{n0324} and \eqref{magneticstabtiliy}.
Using \eqref{uestimeate}--\eqref{heighesimte}, we deduce from \eqref{0106}$_2$ that
\begin{equation}\label{03241}
\begin{aligned}
\mu\|\nabla{ u}\|_{L^2}^2 + \|\sqrt{ \bar{\rho}} u_t \|_{L^2}^2 = & -\int (\lambda_0 m N\cdot\partial_i u+\varrho g u_3)\mm{d}x \\
\leq&  g\|\varrho \|_{L^2} \| { u}\|_{L^2} + \lambda_0 m  \|  N\|_{L^2}\|\partial_i { u}\|_{L^2}\\[1mm]
\leq & C\| (\varrho_0, \partial_i{u}_0,\mu\Delta {u}_0, N_0,\partial_i N_0)\|_{L^2}^2,
\end{aligned}\end{equation}
which yields \eqref{0324}. Hence, to complete the proof of the linear stability in Theorem \ref{thm:0201},
it remains to show the asymptotic stability of $(\varrho,u, N)$ in \eqref{instabi}.

By \cite[Theorem 1.68]{NASII04}, Sobolev's inequality, the estimates on $\|u\|_{W^{1,2}(\mathbb{R}^+,H^1)}$ in \eqref{uestimeate} and \eqref{heighesimte}, we infer that
\begin{equation*}\label{convergence}
\begin{aligned}\int_0^t\left|\frac{\mm{d}}{\mm{d}\tau}\|( {u},\nabla {u}) \|_{L^2}^2\right|\mm{d}\tau
 &\leq 2\int_0^t\left(\| {u}\|_{L^2}\| {u}_\tau\|_{H^{-1}} + \|\nabla {u}\|_{L^2}\|\nabla {u}_\tau\|_{H^{-1}}\right)\mm{d}\tau \\
&\leq C\int_0^t \|( {u},\nabla {u}, {u}_\tau,\nabla {u}_\tau)\|_{L^2}^2\mm{d}\tau  \\
&\leq C\int_0^t \| ( {u},{u}_\tau)\|_{H^1}^2\mm{d}\tau\leq  C\| (\varrho_0,   \partial_i{u}_0,\mu\Delta u_0, \partial_iN_0)\|_{L^2}^2.
\end{aligned}\end{equation*}
Hence, $\| {u}(t)\|_{H^1}\in W^{1,1}(0,\infty)$, which, together with \eqref{03241},  implies   $\| {u}(t)\|_{H^1}$ and $\|u_t(t)\|_{L^2}\to 0$ as $t\to\infty$.
Using \eqref{n0324} and \eqref{magneticstabtiliy}, we see that there are two measurable
functions $\varrho_\infty$ and $N_\infty$ and a time sequence $\{t_n\}_{n=1}^\infty\subset\mathbb{R}^+$, such that
$$(\varrho,N)(t_n)\rightarrow (\rho_\infty,N_\infty)\;\;\mbox{ weakly in }L^2\mbox{ as }n\rightarrow \infty.$$
On the other hand, we have that for any $t\in \mathbb{R}^+$,
\begin{equation}\label{behaveioruofrho}
\begin{aligned}
\|\varrho(t)-\varrho_\infty\|_{L^2}\leq
&\liminf_{n\rightarrow \infty} \|\varrho(t)-\varrho(t_n)\|_{L^2}
\leq\liminf_{n\rightarrow \infty} \int_t^{t_n}\|\varrho_\tau\|_{L^2}\mm{d}\tau
\\ \leq &\|\bar{\rho}'\|_{L^\infty} \int_t^\infty\|u_3\|_{L^2}\mm{d}\tau\leq C\|  {u}_0\|_{L^2}
\end{aligned}
\end{equation}
and
$$\|N(t)-N_\infty\|_{L^2}\leq \int_t^\infty\|N_\tau\|_{L^2}\mm{d}\tau
\leq m\int_t^\infty\|\partial_i u(\tau)\|_{L^2}\mm{d}\tau\leq C\|  {u}_0\|_{L^2}. $$
Thus, $\|(\varrho-\varrho_\infty,   N-N_\infty)(t)\|_{L^2}\rightarrow 0$ as $t\rightarrow \infty$.
Finally, multiplying \eqref{0106} by $u$, we obtain
\begin{equation*}
- \int  \bar{\rho} u_t\cdot \varphi  \dx=\int( \mu\nabla { u}:\nabla \varphi+
  \lambda_0 m N\cdot \partial_{i}\varphi+\varrho ge_3) \dx,\end{equation*}
  which, together with \eqref{instabi}, implies \eqref{steadystate}.
The proof of the linear stability in Theorem \ref{thm:0201} is complete.

\section{Proof of Theorem \ref{thm:0202}}\label{sect:04}
In this section, we adapt the basic ideas in the proof of Theorem \ref{thm:0201} to show Theorem \ref{thm:0202}. Due to the compressibility,
the proof of Theorem \ref{thm:0202} will be more complicated than that of Theorem \ref{thm:0201}.

\subsection{Linear instability}\label{sect:0401}
We still apply a modified variational method to construct a solutions of the boundary value problem \eqref{com0201},
so we modify \eqref{com0201} as follows.
 \begin{equation}\label{com0401}
\left\{  \begin{array}{l} \alpha \bar{\rho}\tilde{u} =s\mu\Delta\tilde{ {{u}}}
 +s\mu_0\nabla\mm{div}{\tilde{u}}+g\bar{\rho}\mm{div}\tilde{u} e_3+g\bar{\rho}'\tilde{u}_3e_3
+\nabla [p'(\bar{\rho})\mm{div}  ( \bar{\rho}\tilde{u})]  \\[1mm]
   \qquad\qquad +  \lambda_0 m_{\mm{c}}(m_\mm{c}\partial_1^2 \tilde{u} -\bar{M}_\mm{c}\mm{div}\partial_1
\tilde{u}) +\nabla [ \lambda_0m_\mm{c} (m_\mm{c}\partial_2 \tilde{u}_2+m_\mm{c}\partial_3 \tilde{u}_3+m'_\mm{c} \tilde{u}_3) ], \\ [1mm]
\tilde{u}|_{\partial\Omega}= {0},   \end{array}
\right.   \end{equation}
where $\alpha:= \alpha(s)$ depends on $s$. Then, the standard energy functional for  \eqref{com0401}  is given by
\begin{equation*}\label{c0204} \mathcal{E}_{\mm{c}}(\tilde{u},s) =
E_c(\tilde{u}) -s V (\tilde{u})\end{equation*}
with an associated admissible set $\mathcal{A}$. Thus, we find an $\alpha$ by maximizing
\begin{equation}\label{com0406}
\alpha :=\sup_{w\in\mathcal{A} }\mathcal{E}_{\mm{c}}(w,s).\end{equation}
Obviously, $\sup_{w\in\mathcal{A}} \mathcal{E}_{\mm{c}}(w,s)<\infty$ for any $s>0$. Next we show the existence of a maximizer for \eqref{com0406}.

\begin{pro}\label{pro:0401}
 Assume that the   density profile $\bar{\rho}$ satisfies the first two conditions in
\eqref{0102},
then for any but fixed $s>0$, the following assertions are valid.
 \begin{enumerate}
    \item[(1)] $\mathcal{E}_{\mm{c}}(w,s)$ achieves its supremum on $\mathcal{A} $.
   \item[(2)] Let $\tilde{ {u}} $ be a maximizer and $\alpha  $ defined by \eqref{com0406},
then $\tilde{u} \in H^1_0$ is a weak solution to the boundary value problem  \eqref{com0401}.
\item[(3)]
 If $\alpha >0$, then the maximizer $\tilde{ {u}} $ further satisfies
  \begin{eqnarray}\label{qh0208}  &&
\tilde{u}_3\not= 0,\ m_{\mm{c}}\partial_1\tilde{u}_3\not= 0, \\
&& \label{qh020812} |m'_{\mm{c}} \tilde{u}_3+m_{\mm{c}}(\partial_2\tilde{u}_2+\partial_3 \tilde{u}_3)|^2+|m_{\mm{c}}\partial_1\tilde{u}_2|^2\not= 0, \\
&& \label{qh020813}  \tilde{ u}_{1}^2+\tilde{u}_{2}^2\not= 0,
\end{eqnarray}
  where $\tilde{u}_{i}$ denotes the $i$-th component of $\tilde{u}$. In addition,
   \begin{equation}\label{qh0209}
  \mm{div}(\bar{\rho}\tilde{ u}) \not=0, \;\quad\mbox{ provided }\;\bar{\rho}'\geq 0.
  \end{equation}   \end{enumerate}
\end{pro}
\begin{pf}
(1) Let $\{w_n\}_{n=1}^\infty\subset \mathcal{A}$ be a maximizing sequence. Using Cauchy-Schwarz's inequality, one sees that
$$ \begin{aligned}
 \mathcal{E}_{\mm{c}}(w_n,s)=  &E_\mathrm{c}(w_n ) -s V (w_n)\\
\leq &\int g\bar{\rho}'|w_{3}^n|^2\mm{d} x +\int \left(2 g\bar{\rho}\mm{div}w_n  w_{3}^n-p'(\bar{\rho})\bar{\rho}|\mathrm{div}w_n|^2
\right) \dx  -s\mu \int |\nabla w_n|^2 \mm{d} x\\
\leq & \int g\bar{\rho}'|w_{3}^n|^2\mm{d} x +\int \frac{g^2\bar{\rho}|w_{3}^n|^2}{p'(\bar{\rho})}   \dx -s\mu
\int |\nabla w_{n}|^2  \mm{d} x ,  \end{aligned}$$
which yields
$$\begin{aligned}
 s\mu \int |\nabla w_{n}|^2  \mm{d} x  +\mathcal{E}_{\mm{c}}(w_n,s) \leq\int g\bar{\rho}'|w_{3}^n|^2\mm{d} x
+\int \frac{g^2\bar{\rho}|w_{3}^n|^2}{p'(\bar{\rho})} \dx.
\end{aligned}$$
We easily see from the above inequality that $\mathcal{E}_{\mm{c}}(w_n,s)$ is bounded, and consequently, $w_n$ is bounded in $H^1_0$.
So, there exists a $\tilde{w}\in\mathcal{A}$ and a subsequence (still denoted by $w_n$ for simplicity), such that
$w_n\rightarrow \tilde{w} $ weakly in $H^1_0$ and strongly in $L^2$. Moreover, by the lower semi-continuity, one has
\begin{equation*}
\begin{aligned}
\sup_{w\in \mathcal{A}}\mathcal{E}_{\mm{c}}(w,s)= &\limsup_{n\rightarrow
\infty}\mathcal{E}_{\mm{c}}({ w}_n,s)  \\
= &\lim_{n\rightarrow \infty} \int \left(g\bar{\rho}'|w_{3}^n|^2\mm{d} x+2 g\bar{\rho}\mm{div}w_n  w_{3}^n\right)\mm{d}x
-\liminf_{n\rightarrow \infty}\int \left[p'(\bar{\rho})\bar{\rho}|\mathrm{div}w_n|^2 \right.\\
&\left. + \lambda_0m^2_{\mm{c}}(|\partial_{1}w_{2}^n |^2+|\partial_{1}w_{3}^n |^2+ |\partial_{2}w_{2}^n+\partial_{3}w_{3}^n|^2)
 + s  (\mu|\nabla w_n|^2+\mu_0|\mm{div}w_n|^2)\right]\mm{d} x\\
\leq & \mathcal{E}_{\mm{c}}(\tilde{w} ,s)\leq \sup_{w\in\mathcal{A} }\mathcal{E}_{\mm{c}} (w,s),
\end{aligned}\end{equation*}
which shows that $\mathcal{E}_{\mm{c}}(w,s)$ achieves its supremum on $\mathcal{A} $.

(2) To show the second assertion, we write \eqref{com0406} as follows
  \begin{equation}\label{0227}
\alpha=\sup_{ { w}\in {H}^1_0}\frac{\mathcal{E}_\mm{c}(w,s)}{J({ {w}})}.
\end{equation}
For any $\tau\in \mathbb{R}$ and $ \psi\in {H}^1_0$, we take $\tilde{ v}(\tau):=\tilde{ {u}} +\tau\psi$. Then, \eqref{0227} implies
  \begin{equation*}E_\mm{c} (\tilde{v}(\tau),s)-\lambda^2J(\tilde{v}(\tau))\leq 0.
\end{equation*}
Let $I(\tau)=E_\mm{c}(\tilde{v}(\tau),s)-\Lambda^2J(\tilde{ v}(\tau))$,
then
$$\frac{\mm{d} I(\tau)}{\mm{d}\tau}\bigg|_{\tau=0}=0.$$ Hence, a direct computation leads to
  \begin{equation*}\begin{aligned}
\alpha\int \bar{\rho}\tilde{u}\cdot \psi\dx = &\int[
(g\bar{\rho}'\tilde{u}_3e_3+g\bar{\rho}\mm{div}\tilde{u} e_3 )\cdot   \psi+g\bar{\rho}\mm{div}\tilde{\psi} e_3\cdot\tilde{u}]\dx \\
& -\int\{  \lambda_0m^2_{\mm{c}} [\partial_1 \tilde{u}_2\partial_1\psi_2+ \partial_1 \tilde{u}_3 \partial_1\psi_3+
(\partial_2\tilde{u}_2+\partial_3\tilde{u}_3)(\partial_2\psi_2 + \partial_3\tilde{u}_3)] \\
& \qquad\quad +p'(\bar{\rho}) \bar{\rho}\mm{div} \tilde{u}\mm{div}
\psi \}\dx-s\int (\mu\nabla\tilde{u}:\nabla\psi +\mu_0\mm{div}{\tilde{u}}\mm{div}{\psi}) \dx.\end{aligned}  \end{equation*}
Using the condition \eqref{comsteady}, we change the above weak form as follows
  \begin{equation}\label{weakform}\begin{aligned}
\alpha\int \bar{\rho}\tilde{u}\cdot \psi\dx
=&
 -\int[(s\mu_0
+   p'(\bar{\rho})\bar{\rho})\mm{div}
  \tilde{u} +  \lambda_0m^2_{\mm{c}} ( \partial_2 \tilde{u}_2+ \partial_3 \tilde{u}_3 ) ]\mm{div}\psi\dx
\\
   \qquad\qquad&-s\mu\int \nabla \tilde{ {{u}}}:\nabla \psi\dx+\lambda_0  m^2_{\mm{c}}\int
(
\mm{div}\tilde{u} \partial_1 \psi_1- \partial_1  \tilde{u} \cdot\partial_1 \psi)\dx\\
&+\int
[(g\bar{\rho}'\tilde{u}_3e_3 +g\bar{\rho}\mm{div}\tilde{u} e_3+ \nabla(  p'(\bar{\rho})
   \bar{\rho}'\tilde{u}_3) + \lambda_0 \nabla (m_{\mm{c}} m'_{\mm{c}}
\tilde{u}_3) ]\cdot \psi\dx,\end{aligned}  \end{equation}
 which implies that $\tilde{ {u}} $ is a weak solution to \eqref{com0401}.

(3) Next, we turn to the proof of \eqref{qh0208}--\eqref{qh0209} by contradiction.

By the second assertion, we know the maximizer $\tilde{ {u}}\in \mathcal{A} $ satisfies \eqref{weakform}, thus $\alpha=\mathcal{E}_{\mm{c}}(\tilde{u},s)$.
Suppose  $\tilde{ {u}}_3= 0$, then $\alpha=\mathcal{E}_{\mm{c}}(\tilde{u},s)<0$ due to  $\int\bar{\rho}|\tilde{u}|^2\mm{d}x=1$,
which contradict with the condition $\alpha>0$. Hence $\tilde{u}_3\not= 0$.

Suppose $|m'_{\mm{c}} \tilde{u}_3+m_{\mm{c}}(\partial_2\tilde{u}_2+\partial_3 \tilde{u}_3)|^2+|m_{\mm{c}}\partial_1\tilde{u}_2|^2 = 0$, then
\begin{equation}\label{muddsf} m'_{\mm{c}} \tilde{u}_3+m_{\mm{c}}(\partial_2\tilde{u}_2+\partial_3 \tilde{u}_3) = 0
\end{equation} and  $m_{\mm{c}}\partial_1\tilde{u}_2 = 0$. Since $m_{\mm{c}}>0$, we get $\partial_1\tilde{u}_2= 0$.
Recalling that $\tilde{u}_2|_{\partial\Omega}=0$, we obtain $\tilde{u}_2= 0$ by a Lagrangian formula \cite[Section 1.3.5.1]{NASII04},
which, tougher with \eqref{muddsf}, implies $\partial_3 (m_{\mm{c}}\tilde{u}_3) =0$. We immediately see
 $\tilde{u}_3\equiv 0$, which contradicts with $\tilde{u}_3\not= 0$. Hence \eqref{qh020812} holds.
Similarly, we can show $m\partial_1\tilde{u}_3\not= 0$.

Suppose  that $\tilde{ u}_{1}^2+\tilde{u}_{2}^2= 0$ or $\mm{div}(\bar{\rho}\tilde{u})= 0$, then
\begin{equation*}\label{horizve}\begin{aligned}0<\alpha=&\int\Big\{g\bar{\rho}'\tilde{{u}}_{3}^2
+ 2g\bar{\rho}\tilde{u}_{3}\partial_{3}\tilde{u}_{3}-
(\bar{p}'(\bar{\rho})\bar{\rho}+
\lambda_0 m^2_{\mm{c}})|\partial_{3}\tilde{u}_{3}|^2-\lambda_0 m^2_{\mm{c}}|\partial_{1}\tilde{u}_3 |^2\Big\}\mm{d}  x\\
&-s\int\left(\mu|\nabla
\tilde{ u}_{3}|^2+\mu_0  |\partial_{3}\tilde{u}_{3}|^2\right)\mm{d}x \\
=&-\int\big[ (\bar{p}'(\bar{\rho})\bar{\rho}+\lambda_0 m^2_{\mm{c}})|\partial_{3}\tilde{u}_{3}|^2+
\lambda_0 m_{\mm{c}}^2|\partial_{1}\tilde{u}_3|^2\big]\mm{d}x\\
&-s\int\left(\mu|\nabla
\tilde{ u}_{3}|^2+\mu_0  |\partial_{3}\tilde{u}_{3}|^2\right)\mm{d}x < 0,
\end{aligned}
\end{equation*}
or
\begin{equation*}\begin{aligned} \label{densityproe}
0<\alpha=&\int\Big\{g\bar{\rho}'\tilde{u}_{3}^2 +(2g\bar{\rho}\tilde{{u}}_{3}-\bar{p}'(\bar{\rho})\bar{\rho}\mm{div}
\tilde{u})\mm{div}\tilde{u}\\
&-\lambda_0m^2_{\mm{c}}(|\partial_{1}\tilde{u}_2 |^2+|\partial_{1}\tilde{u}_3 |^2
+|\partial_{2}\tilde{u}_2+\partial_3\tilde{u}_3|^2 \Big\}\mm{d}x -s\int\left(\mu|\nabla
\tilde{ u}|^2+\mu_0 |\mm{div}\tilde{u}|^2\right)\mm{d} x\\
=&-\int\big[ g\bar{\rho}'\tilde{{u}}_{3}^2
+\bar{p}'(\bar{\rho})\bar{\rho}|\mm{div}\tilde{u}|^2+\lambda_0m^2_{\mm{c}}(|\partial_{1}
\tilde{u}_2 |^2+|\partial_{1}\tilde{u}_3 |^2
+|\partial_{2}\tilde{u}_2+\partial_3\tilde{u}_3|^2\big] \mm{d}x\\
& -s\int\left(\mu|\nabla\tilde{u}|^2+\mu_0 |\mm{div}\tilde{u}|^2\right)\mm{d} x<0 ,
\end{aligned}\end{equation*}
which is a contradiction. Therefore, \eqref{qh020813} and \eqref{qh0209} hold.
 This completes the proof.
\hfill $\Box$
\end{pf}
\begin{pro}\label{pro:0402} Let $\tilde{u} \in H^1_0$
be a weak solution of the boundary value problem \eqref{com0401}, then $\tilde{u}\in H^2$.
\end{pro}
\begin{pf} Firstly, we write \eqref{com0401} as follows.
\begin{equation}\label{com0401newform}
\left\{   \begin{array}{l}  (s\mu+s\mu_0+p'(\bar{\rho})
 \bar{\rho})\partial_1^2\tilde{ {{u}}}_1+s\mu\partial_2^2\tilde{ {{u}}}_1 +s\mu\partial_3^2\tilde{ {{u}}}_1  \\
\;\; + (s\mu_0+p'(\bar{\rho}) \bar{\rho})\partial_1\partial_2 \tilde{u}_2 + (s\mu_0+p'(\bar{\rho})
    \bar{\rho})\partial_1\partial_3 \tilde{u}_3 =\alpha\bar{\rho}\tilde{u}_1+g\bar{\rho}\partial_1 \tilde{u}_3,\\[1mm]
(s\mu + \lambda_0 m^2_{\mm{c}})\partial_1^2\tilde{ {{u}}}_2+(s\mu+s\mu_0 + p'(\bar{\rho})\bar{\rho}
+\lambda_0m^2_{\mm{c}})\partial_2^2\tilde{ {{u}}}_2 + s\mu\partial_3^2\tilde{ {{u}}}_2 \\
\;\; +(s\mu_0 +p'(\bar{\rho})\bar{\rho})\partial_1\partial_2{\tilde{u}}_1   +(s\mu_0+\lambda_0m^2_{\mm{c}}+
p'(\bar{\rho})\bar{\rho} )\partial_2\partial_3{\tilde{u}}_3= \alpha\bar{\rho}\tilde{u}_2+g\bar{\rho}\partial_2 \tilde{u}_3, \\[1mm]
(s\mu+\lambda_0 m^2_{\mm{c}})\partial_1^2\tilde{ {{u}}}_3+s\mu\partial_2^2\tilde{ {{u}}}_3+\partial_3[(s\mu_0+ p'(\bar{\rho})\bar{\rho}
  )\partial_1 \tilde{u}_1 \\
\;\; +(s\mu_0+\lambda_0m^2_{\mm{c}} +p'(\bar{\rho})\bar{\rho} )  \partial_2 \tilde{u}_2
+ (s\mu+s\mu_0+\lambda_0m^2_{\mm{c}} +p'(\bar{\rho})\bar{\rho}  )\partial_3 \tilde{ {{u}}}_3 ]\\
= \alpha\bar{\rho}\tilde{u}_3-g\mathrm{div}(\bar{\rho}\tilde{u})+\partial_3( g\bar{\rho}\tilde{u}_3),\\[1mm]
\tilde{u}|_{\partial\Omega}= {0},
  \end{array}
\right. \end{equation}
Let $f=(\alpha\bar{\rho}\tilde{u}_1+g\bar{\rho}\partial_1 \tilde{u}_3,\alpha\bar{\rho}\tilde{u}_2+g\bar{\rho}\partial_2 \tilde{u}_3, \alpha\bar{\rho}\tilde{u}_3-g\mathrm{div}(
\bar{\rho}\tilde{u})+\partial_3( g\bar{\rho}\tilde{u}_3))^{T}$ and $(A_{ij}^{\alpha\beta})_{1\leq i, j\leq 3}^{1\leq \alpha,\beta\leq 3}$ be the    matrix  of coefficients of the linear
elliptic equations \eqref{com0401newform}, then \eqref{com0401newform} can be written as $$-\partial_\alpha
(A_{ij}^{\alpha\beta}\partial_\beta u_j)=f_i,$$
where we have used the Einstein convention of summing over repeated indices, and  the non-zero coefficients are
$$
\begin{aligned}
&A_{11}^{22}=A_{11}^{33}=A_{22}^{33}= A_{33}^{22}=s\mu,\\
& A_{12}^{12}=A_{13}^{13}=A_{21}^{21}=A_{31}^{31}= s\mu_0+p'(\bar{\rho}) \bar{\rho},\\
& A_{22}^{11}=A_{33}^{11}=s\mu+\lambda_0 m^2_{\mm{c}},\quad A_{32}^{32}=A_{23}^{23}= s\mu_0+p'(\bar{\rho})\bar{\rho}
+\lambda_0 m^2_{\mm{c}},
\\& A_{11}^{11}=s\mu+s\mu_0+p'(\bar{\rho}) \bar{\rho},\quad A_{22}^{22}=s\mu+s\mu_0+
p'(\bar{\rho})\bar{\rho}+\lambda_0m^2_{\mm{c}},\\
& A_{33}^{33}=s\mu+s\mu_0 +p'(\bar{\rho})\bar{\rho}+\lambda_0m^2_{\mm{c}}.
\end{aligned}$$
Noting that, for any $\xi$, $\eta\in \mathbb{R}^3$,
$$\begin{aligned}A_{ij}^{\alpha\beta}\xi_\alpha\xi_\beta\eta_i\eta_j=&s\mu|\xi|^2|\eta|^2+
(s\mu+p'(\bar{\rho})\bar{\rho})(\xi_1\eta_1+\xi_2\eta_2
+\xi_3\eta_3)^2\\&+\lambda_0 m^2_{\mm{c}}(\xi_2\eta_2+\xi_3\eta_3)^2+\lambda_0 m^2_{\mm{c}}\xi_1^2(\eta_2^2+\eta_3^2)\geq s\mu|\xi|^2|\eta|^2,
\end{aligned}$$
hence $A_{ij}^{\alpha\beta}$ satisfies the strong elliptic condition. On the other hand, $\partial\Omega$ is of class $C^{2}$,  $A_{ij}^{\alpha\beta}\in C^{0,1}(\Omega)$ and
$f\in L^{2}$. Thus, if we apply \cite[Theorem 4.11]{AnintroudctuionGMML} to the weak form \eqref{weakform}, we get $\tilde{u}\in H^2$.\hfill$\Box$
\end{pf}

Similarly to Proposition \ref{pro:0202}, we have the following properties of $\alpha(s)$ as a function of $s> 0$.
\begin{pro}\label{pro:c0202}
Under the assumptions of Proposition \ref{pro:0401}, the function $\alpha(s)$ defined on $(0,\infty)$ enjoys the following properties:
\begin{enumerate}[\quad \ (1)] \item $\alpha(s)\in C_{\mathrm{loc}}^{0,1}(0,\infty)$ is nonincreasing.
   \item If $\mm{C_r}>0$, then there are constants $c_6$, $c_7>0$ which depend on $g$, $\bar{\rho}$, $\mu$, $\mu_0$, $\lambda_0$, $p$ and $m_{\mm{c}}$, such that
  \begin{equation*}  \alpha(s)\geq c_6-sc_7. \end{equation*}
  \end{enumerate}
\end{pro}
\begin{pf}
The monotonicity
and the second assert obviously hold by directly following the proof of Proposition \ref{pro:0202}, while the absolute continuity
  can be established by modifying the proof of the first assertion in Proposition \ref{pro:0202}.
  Here we give the proof of absolute continuity for the reader's convenience.

Let $I:=[a,b]\subset (0,\infty)$ be a bounded interval. For any $w\in \mathcal{A}$, by Cauchy-Schwarz's inequality,
\begin{equation*}  \begin{aligned}
 \mathcal{E}_{\mm{c}}(w,s) \leq &\int (g\bar{\rho}'w_3^2
+2g\bar{\rho}\mm{div}w w_3)\mm{d}  x-  \int\bar{p}'(\bar{\rho})\bar{\rho}|\mm{div}w|^2\mm{d} x  \\
\leq &{g}\left(\left\|\frac{\bar{\rho}'}{\bar{\rho}}
\right\|_{L^\infty}+ \left\|\frac{  g\bar{\rho}}{\bar{p}'(\bar{\rho}) }\right\|_{L^\infty}\right).
\end{aligned}\end{equation*}
Hence, from the monotonicity of $\alpha (s)$ we get
\begin{equation} \label{nc11}
|\alpha(s)|\leq \max\left\{|\alpha(a)|,{g}\left(\left\|\frac{\bar{\rho}'}{\bar{\rho}}
\right\|_{L^\infty}+ \left\|\frac{g\bar{\rho}}{\bar{p}'(\bar{\rho}) }\right\|_{L^\infty}\right)\right\}:=
L<\infty\quad\mbox{ for any }s\in I.
\end{equation}
On the other hand, for any $s\in I$, there exists a maximizing sequence
$\{w_n^{s}\}\subset\mathcal{A}$ of $\sup_{w\in
\mathcal{A}}\mathcal{E}_{\mm{c}}(w,s)$, such that
\begin{equation}\label{0232n}\begin{aligned}|\alpha(s)- \mathcal{E}_{\mm{c}}(w^{s}_n,s)|<1
\end{aligned}.\end{equation}
Making use of  \eqref{nc11} and \eqref{0232n}, we infer from the definition of $\mathcal{E}_{\mm{c}}(w,s)$ that
\begin{equation*}\begin{aligned}
0\leq & \int\left(\mu |\nabla w|^2+\mu_0 |\mm{div} w|^2\right)\mm{d}x \\
\leq &\frac{1}{s}\int  [g\bar{\rho}'|w^{s}_{n3}|^2 + (2g\bar{\rho}w_{n3}^s
-\bar{p}'(\bar{\rho})\bar{\rho}\mm{div}w^s_n)\mm{div}w^s_n]\mm{d}x -\frac{\mathcal{E}_{\mm{c}}(w^s_n,s)}{s} \\
\leq & \frac{1+L}{a}+\frac{g}{a}\left[\left\|\frac{\bar{\rho}'}{\bar{\rho}}
\right\|_{L^\infty}+ \left\|\frac{g\bar{\rho}}{\bar{p}'(\bar{\rho}) }\right\|_{L^\infty}\right] \leq \frac{1+2L}{a}.
\end{aligned}\end{equation*}
Thus, for $s_i\in I$ ($i=1,2$), we have
\begin{equation*}\begin{aligned} \alpha(s_1)
\leq  \alpha(s_2)+(1+2L)|s_1-s_2|/a,
\end{aligned}\end{equation*}
and further infer that
\begin{equation*}\begin{aligned}|\alpha(s_1)-\alpha(s_2)|\leq (1+2L)|s_1-s_2|/a,
\end{aligned}\end{equation*}
which yields $\alpha(s)\in C_{\mathrm{loc}}^{0,1}(0,\infty)$.
 This completes the proof of Proposition \ref{pro:c0202}.
 \hfill $\Box$
\end{pf}

With Proposition \ref{pro:c0202} in hand, we can directly follow the proof of \eqref{growth} to deduce that there is a unique $\Lambda>0$ satisfying
\begin{equation} \label{growth123}
 \Lambda=\sqrt{\alpha(\Lambda)}=\sqrt{\sup_{ { w}\in
\mathcal{A}}\mathcal{E}_{\mm{c}}(w,\Lambda)}>0. \end{equation}
Thus, by virtue of Propositions \ref{pro:0401} and \ref{pro:0402}, there is a solution $\tilde{u}\in H^2$
to the boundary value problem \eqref{com0201} with $\Lambda$ constructed in \eqref{growth123},
where $\tilde{u}$ satisfies \eqref{qh0208}--\eqref{qh0209}. Thus, we conclude the following proposition, which gives
the linear instability in Theorem \ref{thm:0202}.
\begin{pro}\label{pro:nnn0203com}
 Assume that the density profile $\bar{\rho}$ satisfies
\eqref{0102} and $\mm{C_r}>0$. Then there exists a $\tilde{u}\in H^2\cap\mathcal{A}$ which solves the boundary value problem \eqref{com0201}
with a finite growth rate $\Lambda>0$ satisfying \eqref{0111nn}. Moreover, $\tilde{u}$ satisfies \eqref{qh0208}--\eqref{qh0209}.
 In particular, let $(\tilde{\rho},\tilde{N}):=(-\mm{div}(\bar{\rho}\tilde{u}),m_{\mm{c}}\partial_1\tilde{u}
 -\tilde{u}_3 \bar{M}_{\mm{c}}'-\bar{M}_{\mm{c}}\mm{div}\tilde{u})/\Lambda$, then
 $(\tilde{\rho},\tilde{u},\tilde{N})\in L^2\times H^2\times L^2$ solves \eqref{comlinear},
$\tilde{N}^2_1+\tilde{N}^2_2 \not= 0$ and $\tilde{N}_3\not= 0$.
In addition, $\tilde{\rho}\not= 0$ provided $\bar{\rho}'\geq 0$.
\end{pro}

\subsection{Linear stability}\label{linearstabofPrar}
Firstly, following the process of the iterative method as in Subsection \ref{stability}, we can show that there exists a unique local-in-time solution
$(\varrho, u, N)$ of the linearized Parker problem \eqref{c0104}--\eqref{lincom}, where we should use the classical regularity
theory on elliptic equations instead of that for the Stokes problem (see \eqref{estimsH2eo28} later).
Moreover, the solution satisfies the following regularity
\begin{equation*}
\begin{aligned}
&\varrho\in C^0(\bar{I}_{{T}^*},H^1),\ (N,\partial_1 N,\nabla N_1)\in   C^0(\bar{I}_{{T}^*},L^2),\\
 &u\in C^0(\bar{I}_{T^*},H^2),\;\; u_t\in C^0(\bar{I}_{T^*},L^2)\cap L^2({I}_{T^*},H^1),
\end{aligned}  \end{equation*}
where $T^*\in (0,1)$ depends on $(\|\varrho_0\|_{H^1}^2+\|u_0\|_{H^2}^2+\|(N_3^0,\partial_1 N_0, \nabla N_1^0)\|_{L^2}^2)$.
Next, we deduce some global-in-time estimates on $\|\varrho(t)\|_{H^1}$, $\|u(t)\|_{H^2}$ and $\|(N_3,\partial_1 N, \nabla N_1)(t)\|_{L^2}^2$.

In view of the regularity of $(\varrho,u, N)$, we deduce from \eqref{newsytem12}$_1$ that for a.e. $t>0$,
\begin{equation*}
\begin{aligned}  & \frac{1}{2}\frac{\mm{d}}{\mm{d}t}\int\bar{\rho}| u_t|^2
\mathrm{d} x= <\bar{\rho}{u}_{tt},{u}_t>  \\
&= \int [(p'(\bar{\rho})\varrho_t+ \lambda_0  m_{\mm{c}} \partial_tN_1)\mm{div}u_t+
  \lambda_0 m_{\mm{c}}' \partial_tN_3\partial_t u_1 \\
&\qquad +  \lambda_0 m_{\mm{c}} \partial_1  N_t\cdot u_t-\varrho_t  g\partial_tu_3] \mm{d}x-\int(\mu
|\nabla u_t|^2+\mu_0|\mm{div}{ u}_t|^2) \mm{d} x:=L_1+L_2,
\end{aligned}\end{equation*}
where $<\cdot,\cdot>$ denotes the  dual product between the spaces $H^{-1}$ and $H_0^1$, and $H^{-1}$ denotes the dual space of $H_0^1$.
On the other hand, using \eqref{lincom}$_1$, \eqref{lincom}$_2$, $\eqref{comsteady}$ and partial integrations, we have
\begin{equation*}
\begin{aligned}
L_1 =&   \int  [-p'(\bar{\rho})\mm{div}( \bar{\rho} {u})+ \lambda_0  m_{\mm{c}} (m_{\mm{c}} \partial_1 u_1
 -m'_{\mm{c}}u_3-m_{\mm{c}}\mm{div}u)]\mm{div}u_t\mm{d}x \\
& \quad  +\int[  \lambda_0 m_{\mm{c}}^2 (  \partial_1^2 u\cdot  u_t  -   \partial_1\mm{div}u \partial_tu_1)
 +g\mm{div}( \bar{\rho} {u}) \partial_t u_3] \mm{d}x \\
=& \int \left\{g\bar{\rho}'u_3\partial_t u_3-(p'(\bar{\rho})\bar{\rho}' +\lambda_0 m_{\mm{c}}m_{\mm{c}}')u_3\mm{div}u_t
+g\bar{\rho}\mm{div}u  \partial_tu_3 - p'(\bar{\rho})\bar{\rho} \mathrm{div}u  \mathrm{div}u_t \right.\\
&\qquad-\lambda_0  m_{\mm{c}}^2[ \partial_1 u\cdot  \partial_1u_t  + ( \mm{div}u -\partial_1 u_1
)\mm{div}u_t - \mm{div}u \partial_t\partial_1 u_1] \mm{d} x\\
=& \int  \{ g\bar{\rho}'u_3\partial_t u_3 +  g\bar{\rho}\partial_t(\mm{div}u  u_3)-  p'(\bar{\rho})\bar{\rho} \mathrm{div}u  \mathrm{div}u_t\\
&\qquad  -  \lambda_0m^2_{\mm{c}}[ \partial_{1}u_2\partial_t\partial_{1} u_2 + \partial_{1}u_3 \partial_t \partial_1 u_3
    +(\partial_{2}u_{2}+ \partial_{3}u_{3}) (\partial_{2}u_{2}+ \partial_{3}u_{3})_t]  \dx \\
=& \frac{1}{2}\frac{\mm{d}}{\mm{d}t}E_\mm{c}(u(t)).
\end{aligned}\end{equation*}
Combining the above two equalities, we get
\begin{equation*}   \begin{aligned}
&\frac{1}{2} \frac{\mm{d}}{\mm{d}t}\left(\int\bar{\rho}| u_t|^2\mathrm{d} x - E_\mm{c}(u(t)) \right) +
\int (\mu|\nabla u_t|^2+\mu_0|\mm{div}u_t|^2)\mm{d} x =0, \end{aligned}
\end{equation*}
which yields
\begin{equation}\label{definforI0}
\begin{aligned}
&\int\bar{\rho}| u_t|^2 \mathrm{d} x - E_\mm{c}(u(t)) +  2\int_0^t\int (\mu|\nabla  u_\tau|^2+
\mu_0|\mm{div}  u_\tau|^2)\mm{d} x\mm{d}\tau  \\
& =\int \Big(\bar{\rho}|u_t|^2 \big|_{t=0}-E_{\mm{c}}(u(0))\Big)\mm{d} x :=I_0.
\end{aligned}\end{equation}
Keeping in mind that $\|\nabla \mm{div}u_0\|_{L^2}\leq\|\Delta u_0\|_{L^2}$ (see \cite[Corollary 9.10]{GDTR}) and
\eqref{poincareinequality}, we easily see that $I_0$ can be bounded from above by \eqref{bondCby}.

Recalling \eqref{comkestep}, we have
\begin{equation*}     \begin{aligned}
& \int\bar{\rho} |u_t|^2\mm{d}x-\lambda_0{\mm{C_r}}
 \int (|\partial_{1} {u}_2 |^2+|\partial_{1}{u}_3 |^2 +|\partial_{2}{u}_2+\partial_3{u}_3|^2) \mm{d}x\\
&+  2\int_0^t\int (\mu|\nabla  u_\tau|^2+ \mu_0|\mm{div} u_\tau|^2)\mm{d} x\mm{d}\tau\leq I_0.
\end{aligned}\end{equation*}
 Since $ {\mm{C_r}}<0$, we obtain by exploiting Poinc\'are's inequality and \eqref{poincareinequality} that
\begin{equation}\label{newesimt05}
\begin{aligned}
&\|u_t(t)\|^2_{L^2}+\|(u_2,u_3,\partial_{1} {u}_2,\partial_{1}{u}_3,\partial_{2}{u}_2+\partial_3{u}_3 )(t)\|^2_{L^2} \\
& + \int_0^t(\mu\|  u_\tau\|^2_{H^1}+
\mu_0\|\mm{div}u_\tau\|^2_{L^2})\mm{d}\tau\leq CI_0,\;\;\mbox{ for any }t\in \mathbb{R}^+.
\end{aligned}\end{equation}
On the other hand,  \eqref{definforI0} can be rewritten as follows
\begin{equation*}  \begin{aligned}
&\int\left[\bar{\rho}| u_t|^2 + p'(\bar{\rho})\bar{\rho}|\mathrm{div} {u}|^2
+ \lambda_0m^2_{\mm{c}}(|\partial_{1} {u}_2 |^2+|\partial_{1} {u}_3 |^2
+|\partial_{2} {u}_2+\partial_3 {u}_3|^2)\right] \mathrm{d} x \\
&+  2\int_0^t\int (\mu|\nabla u_\tau|^2+ \mu_0|\mm{div} u_\tau|^2)\mm{d} x\mm{d}\tau = I_0
+\int \left(2 g\bar{\rho}\mm{div}\tilde{u} {u}_3+g\bar{\rho}' {u}_3^2\right)\mm{d}x.
\end{aligned}\end{equation*}
Thus, using Cauchy-Schwarz's inequality and \eqref{newesimt05}, we obtain
\begin{equation}\label{intiestimate1}
\begin{aligned}
\|\mathrm{div} {u} \|^2_{L^2} \leq CI_0,\end{aligned}\end{equation}
which, together with \eqref{newesimt05} and \eqref{poincareinequality}, yields
\begin{equation}\label{intiestimate2}
\begin{aligned}    \|(u_1,\partial_1 u_1)\|^2_{L^2}\leq CI_0 \end{aligned}\end{equation}
 Putting \eqref{newesimt05}--\eqref{intiestimate2}, \eqref{lincom}$_1$ and \eqref{lincom}$_2$ together,
 we get the stability estimate \eqref{comuestimeate}.

Now, we turn to the estimation of $\|u(t)\|_{H^2}$. To this end, we use  \eqref{lincom}$_1$ and \eqref{lincom}$_3$ to rewrite
\eqref{lincom}$_2$ as follows.
\begin{equation}\label{n0426}
 \begin{aligned}
\mu\Delta{ u}+\mu_0\nabla\mm{div}{ u} =\bar{\rho} u_t +\nabla \left(p'(\bar{\rho})\varrho+ \lambda_0
 m_{\mm{c}}N_1\right)  -  \lambda_0 N_3  \bar{M}'_{\mm{c}} - \lambda_0 m_{\mm{c}} \partial_1 N+ \varrho ge_3,
\end{aligned} \end{equation}
where
$$\varrho = -\int_0^t\mm{div}( \bar{\rho} {u})(\tau)
\mm{d}\tau+\varrho_0\mbox{ and }N =  \int_0^t(m_{\mm{c}} \partial_1 u
 -u_3  \bar{M}_{\mm{c}}'-\bar{M}_{\mm{c}}\mm{div}u)(\tau)\mm{d}\tau+N_0. $$
The viscosity term in \eqref{n0426} defines a strong elliptic operator on $u$, thus
\begin{equation}\label{estimsH2eo28}
\begin{aligned}
 \|   u(t)\|_{H^2} \leq &\|\bar{\rho} u_t +\nabla \left(p'(\bar{\rho})\varrho+ \lambda_0
 m_{\mm{c}}N_1\right)  -
  \lambda_0 N_3  \bar{M}'_{\mm{c}}-
  \lambda_0 m_{\mm{c}} \partial_1 N+ \varrho ge_3\|_{L^2}\\
\leq &C_{\mu,\mu_0}\left(\|\varrho_0\|_{H^1}+\|( N_3^0,\partial_1 N_0,\nabla N_1^0, u_t ) \|_{L^2}+\int_0^t\| u(\tau)\|_{H^2}\mm{d}\tau\right)\\
\leq &C_{\mu,\mu_0}\left(\|\varrho_0\|_{H^1}+\|( N_3^0,\partial_1 N_0,\nabla N_1^0)\|_{L^2}+I_0+\int_0^t\| u(\tau)\|_{H^2}\mm{d}\tau\right) ,
\end{aligned}
\end{equation}
which, together with Grownwall's inequality, gives
\begin{equation}\label{gloableestimac}
\begin{aligned}
 \|   u (t)\|_{H^2}\leq  C_{\mu,\mu_0} (\|\varrho_0\|_{H^1}+\|( N_3^0,\partial_1 N_0,\nabla N_1^0,)\|_{L^2}+I_0)
\left(1+C_{\mu,\mu_0}t e^{C_{\mu,\mu_0}t}\right).
\end{aligned}
\end{equation}
Here $C_{\mu,\mu_0}$ denotes a generic positive constant depending on $\Omega$, $\mu$, $\mu_0$ and the other known physical parameters.
With the help of \eqref{gloableestimac} and
\begin{equation*}
\begin{aligned}
 \|\varrho(t)\|_{H^1}+\|(N_3, \partial_1 N,\nabla N_1)(t)\|_{L^2}\leq  &
C  \left( \|\varrho_0\|_{H^1}+\|(N_3^0, \partial_1 N_0,\nabla N_1^0)\|_{L^2}+\int_0^t\| u(\tau)\|_{H^2}\mm{d}\tau\right),
\end{aligned}
\end{equation*}
we immediately obtain a global solution $(\varrho, u,N)$ by a continuity argument based on the local well-posedness result.
Moreover, the global solution satisfies  the stability estimate \eqref{comuestimeate}.

We proceed to deriving the estimates \eqref{uestimeate2c}--\eqref{0324c}.
Firstly,  we get from \eqref{n0426} that
\begin{equation*}\label{timesdeine1co}  \begin{aligned}
&\bar{\rho} u_t +\nabla \left[ - p'(\bar{\rho})\mm{div}\left(\bar{\rho}\int_0^tu(\tau)\mm{d}\tau\right)+ \lambda_0
 m_{\mm{c}}  \int_0^t (m_{\mm{c}}\partial_1u_1 -m'_{\mm{c}}u_3-m_{\mm{c}}\mm{div}u )(\tau)\mm{d}\tau
\right]\\
&=\mu\Delta{ u}+\mu_0\nabla\mm{div}{ u}+   \lambda_0 m_{\mm{c}} \partial_1  \int_0^t( m_{\mm{c}} \partial_1 u
 -u_3  \bar{M}_{\mm{c}}'-\bar{M}_{\mm{c}}\mm{div}u)(\tau)\mm{d}\tau\\
&\quad + \lambda_0 m_{\mm{c}} \partial_1 \int_0^t u_3(\tau)\mm{d}\tau\bar{M}_{\mm{c}}'+g\mm{div}\left(\bar{\rho}
\int_0^t u(\tau)\mm{d}\tau \right)e_3+P_0.
\end{aligned}
\end{equation*}
Similarly to the identity \eqref{engeryequality}, we have
\begin{equation}\label{definforI0newein}
\begin{aligned}
& \int\bar{\rho}| u |^2 \mathrm{d} x- E_\mm{c}\left(\int_0^tu(\tau)\mm{d}\tau\right)
+  2\int_0^t\int (\mu|\nabla  u |^2+ \mu_0|\mm{div}  u |^2)(\tau)\mm{d} x\mm{d}\tau\\
 &=\int  \bar{\rho}|u_0|^2 \mm{d} x+2\int P_0\int_0^t u(\tau)\mm{d}\tau\mm{d}x.
\end{aligned}\end{equation}
Noting that $\int_0^t u(\tau)\mm{d}\tau\big|_{\partial\Omega}=0$, we obtain \eqref{uestimeate2c}
from \eqref{definforI0newein} by following the derivation of \eqref{comuestimeate}.
Utilizing \eqref{lincom}$_1$, we find that
\begin{equation}\begin{aligned}\label{0323c}
\|\varrho(t)\|_{L^2}\leq &\|\varrho_0\|_{L^2} +\left\|\int_0^t\mm{div}(\bar{\rho}u)(\tau) \mm{d}\tau\right\|_{L^2}\\
=&\|\varrho_0\|_{L^2}+\left\|\mm{div}\left(\bar{\rho}\int_0^t u(\tau)\mm{d}\tau\right) \right\|_{L^2}\leq C(\|(\varrho_0,u_0,P_0)\|_{L^2}+I_0).
\end{aligned}\end{equation}
 Similarly, using \eqref{lincom}$_3$, one obtains
\begin{equation}\label{magneticstabtiliyc}\begin{aligned}
\| N(t)\|_{L^2} \leq&  \| N_0\|_{L^2}+\left\|\int_0^t(m_{\mm{c}} \partial_1 u
 -u_3 \bar{M}_{\mm{c}}'-\bar{M}_{\mm{c}}\mm{div}u)(\tau)\mm{d}\tau\right\|_{L^2}\\
 \leq& C(\|(N_0,u_0,P_0)\|_{L^2}+I_0).
\end{aligned}\end{equation}
Combining \eqref{0323c} with \eqref{magneticstabtiliyc}, we arrive at \eqref{heighesimtec}.

Now, making use of \eqref{uestimeate2c}, \eqref{heighesimtec} and  Cauchy-Schwarz's inequality, we infer from \eqref{lincom}$_2$ that
\begin{equation*}\label{03241c}
\begin{aligned}
&\mu\|\nabla{ u} \|_{L^2}^2+\mu_0\|\mm{div}{ u} \|_{L^2}^2\\
&= -  \int [\bar{\rho}u_t \cdot u  -( p'(\bar{\rho})\varrho+ \lambda_0  m_{\mm{c}} N_1)\mm{div}u-
  \lambda_0 m'_{\mm{c}}N_3u_1 + \lambda_0 m_{\mm{c}} N\cdot\partial_1 u+\varrho g u_3]\mm{d}x\\
&\leq C(\|(\varrho, N_3,u_t)\|_{L^2}\|u\|_{L^2}+\|(\varrho, N_1)\|_{L^2}\|\mm{div}u\|_{L^2}+\|N\|_{L^2}\|\partial_1 u\|_{L^2} )\\
 &\leq C(\| (\varrho_0,u_0,  N_0,P_0 )\|_{L^2}^2+I_0),
\end{aligned}\end{equation*}
which implies \eqref{0324c}.
In addition, following the proof of \eqref{steadystate} and \eqref{instabi}, and using the stability estimates,
we obtain the asymptotic behaviors \eqref{instabic1} and \eqref{instabic}.
The proof of linear stability results in Theorem \ref{thm:0202} is complete.

\setcounter{equation}{0}  

\section{Additional results}\label{sec:05}
\subsection{Critical number of horizontally periodic domains}
In this subsection we prove the equality \eqref{assertirem0202} in Remark \ref{rem0202}. Obviously, it suffices to show
the following conclusion.
\begin{pro}\label{addition1}
Let $L>0$, $l>0$, $\Omega:=(2\pi L\mathbb{T})^2\times (-l,l)$,
$\bar{\rho}:=\bar{\rho}'(x_3)\in L^\infty(\mathbb{R})$, $$a=  \sup_{ { w}(x)\in H^{1}_\sigma} {\frac{ \int\bar{\rho}' { w}_3^2(x)\dx}
 { \int|\partial_3 { w}(x) |^2\dx}}\;\mbox{ and }\; b= {\sup_{\psi(x_3)\in H^1_0(-l,l)}
\frac{ \int_{-l}^l\bar{\rho}'|\psi(x_3)|^2\mm{d}x_3}{ \int_{-l}^l|\psi'(x_3)|^2\mm{d}x_3}}, $$
then $a=b$.
\end{pro}
\begin{pf}
 Let $\hat{w}_3(\xi,x_3)$ be the horizontal Fourier transform of $w_3(x)\in H_\sigma^1$, i.e.,
$$ \hat{w}_3(\xi,x_3)=\int_{(2\pi L\mathbb{T})^2}w_3( x',x_3)e^{-\mm{i} x'\cdot\xi}\mm{d}x', $$
where $x'=(x_1,x_2)$ and $\xi=(\xi_1,\xi_2)$, then $\widehat{\partial_3  w_3} = \partial_{3} \widehat{w}_3$.
We denote $\psi(\xi,x_3):= \psi_1(\xi,x_3) + i\psi_2(\xi,x_3):=\hat{w}_3(\xi,x_3)$, where $\psi_1$ and $\psi_2$ are real functions.
  By the Fubini and Parseval theorems (see \cite[Proposition 3.1.16]{grafakos2008classical}), we have
\begin{equation}\label{0501}\begin{aligned}
\int  \bar{\rho}'|w_3(x)|^2\mm{d} x
=&\frac{1}{4\pi^2 L^2}\sum_{\xi\in (L^{-1}\mathbb{Z})^2}\int_{-l}^l (\bar{\rho}'1_{\{\bar{\rho}'\geq 0\}}
+\bar{\rho}'1_{\{\bar{\rho}'< 0\}})|\hat{w}_3(\xi,x_3)|^2\mm{d}x_3 \\
=&\frac{1}{4\pi^2 L^2}\sum_{\xi\in (L^{-1}\mathbb{Z})^2}\int_{-l}^l \bar{\rho}'(\psi_1^2(\xi,x_3)+\psi_2^2(\xi,x_3))\mm{d}x_3
\end{aligned}\end{equation}and \begin{equation}\label{0502}
\begin{aligned}
\int  |{\partial_3 {w}_3}(x)|^2\mm{d} x= \frac{1}{4\pi^2 L^2}\sum_{\xi\in (L^{-1}\mathbb{Z})^2}\int_{-l}^l
 (|\partial_{3}\psi_1(\xi,x_3)|^2+ |\partial_{3}\psi_2(\xi,x_3)|^2) \mm{d}x_3,
\end{aligned}\end{equation}
where $1_{\{\bar{\rho}'\geq 0\}}$ and $1_{\{\bar{\rho}'< 0\}}$ denote the characteristic functions.
Noting that $\psi_i\in H_0^1(-l,l)$, by the definition of $b$, we have
\begin{equation*}\begin{aligned}
b \displaystyle \int_{-l}^l
 |\partial_{3}\psi_i(\xi,x_3)|^2 \mm{d}x_3
\geq
  \int_{-l}^l\bar{\rho}'|\psi_i(\xi,x_3)|^2\mm{d}x_3,
 \end{aligned}\end{equation*}
Thus, using \eqref{0501}--\eqref{0502}, we immediately deduce that
\begin{equation*}\begin{aligned}
\displaystyle b\geq & \frac{\sum_{\xi\in (L^{-1}\mathbb{Z})^2} \int_{-l}^l\bar{\rho}'|\psi(\xi,x_3)|^2\mm{d}x_3}{\sum_{\xi\in (L^{-1}\mathbb{Z})^2} \int_{-l}^l
 |\partial_{3}\psi(\xi,x_3)|^2 \mm{d}x_3} = \frac{ \int \bar{\rho}'|w_3(x)|^2\mm{d} x}{ \int  |{\partial_3 {w}_3}(x)|^2\mm{d} x}
\geq  \frac{ \int \bar{\rho}'|w_3(x)|^2\mm{d} x}{ \int  |{\partial_3 {w}}(x)|^2\mm{d} x},
\end{aligned}\end{equation*}
Hence $a\leq b$.

Next we turn to the proof of $a\geq b$. We choose a maximizing sequence $\{\psi_j\}_{j=1}^\infty\subset H_0^1(-l,l)$ of $b$ to see at
\begin{equation}\label{badefient}
b= \lim_{j\rightarrow \infty}\frac{ \int_{-l}^l\bar{\rho}'
|\psi_j(x_3)|^2\mm{d}x_3}{ \int_{-l}^l|\psi'_j(x_3)|^2\mm{d}x_3} .
\end{equation}
For each $\psi_j$, we can construct an approximate sequence $\{\psi_{m}\}_{m=1}^\infty\subset C_0^\infty(-l,l)$ satisfying
 $\psi_{m}\rightarrow \psi_j$ in $H_0^1(-l,l)$ as $m\rightarrow \infty$. Now, we denote
 $$\tilde{w}=(0, \psi'_m (x_3)\cos(n L^{-1}x_2 ), n L^{-1}\psi_m(x_3) \sin (nL^{-1} x_2 )),$$ then $\tilde{w}(x)\in H^1_\sigma$ and
$$\begin{aligned}
& {\frac{\int\bar{\rho}' \tilde{ w}_3^2(x)\mm{d}x_3} {\int|\partial_3 \tilde{ w}(x) |^2\dx}}
= {\frac{\int\bar{\rho}' \tilde{ w}^2_3(x)\dx}{\int (|\partial_3 \tilde{ w}_2(x)|^2+|\partial_3 \tilde{ w}_3(x)|^2)\dx}}\\
&= \frac{ (n L^{-1})^2\int_0^{2\pi L} \int_0^{2\pi L} \int_{-l}^l\bar{\rho}'
\psi^2_m(x_3) \sin^2 (n L^{-1} x_2 )\mm{d}x_1\mm{d}x_2\mm{d}x_3}{\int_0^{2\pi L} \int_0^{2\pi L}
\int_{-l}^l( |\psi''_m (x_3)\cos(n L^{-1} x_2)|^2 +|n L^{-1}  \psi'_m (x_3)\sin (n L^{-1} x_2)|^2)\mm{d}x_1\mm{d}x_2\mm{d}x_3}\\
& = \frac{ \int_{-l}^l\bar{\rho}' \psi^2_m(x_3)  \mm{d}x_3}{ L^2 n ^{-2} \int_{-l}^l  |\psi''_m (x_3)|^2   \mm{d}x_3
+ \int_{-l}^l | \psi'_m (x_3) |^2  \mm{d}x_3}\qquad\mbox{ for sufficiently large }m,
\end{aligned} $$
whence,
$$a\geq \lim_{m\rightarrow \infty} \lim_{n\rightarrow \infty}\frac{ \int_{-l}^l\bar{\rho}'
\psi^2_m(x_3)  \mm{d}x_3}{ L^2 n ^{-2} \int_{-l}^l  |\psi''_m(x_3) |^2  \mm{d}x_3 + \int_{-l}^l | \psi'_m(x_3)  |^2  \mm{d}x_3}
= \frac{ \int_{-l}^l\bar{\rho}' \psi^2_j(x_3)  \mm{d}x_3}{   \int_{-l}^l | \psi'_j(x_3)  |^2  \mm{d}x_3}, $$
which, together with \eqref{badefient}, yields $a\geq b$. The proof is complete.\hfill $\Box$
\end{pf}

\subsection{Sharp growth rate of solutions to the linearized problems}\label{sec:04}
In this section we show that $\Lambda$ defined by \eqref{Lambdard} resp. \eqref{0111nn} is the sharp growth rate for any
solution of the linearized problem \eqref{0104}--\eqref{0106} resp. \eqref{c0104}--\eqref{lincom}.
We shall exploit the energy estimates as in \cite{JFJSO2014,HHJGY,GYTI1} to show that $e^{\Lambda t}$
is indeed the sharp growth rate for $(\varrho,u,N)$ in $L^2\times H^1\times L^2$-norm.

 \begin{pro}\label{pro:0303n}
(i) Let $(\varrho, {u}, N )$ solve the linearized  magnetic RT problem \eqref{0104}--\eqref{0106}
with an associated pressure $q$. Then for any $t\geq 0$,
\begin{eqnarray*}
&& \|\varrho(t)\|_{X}^2\leq C_\mu e^{2\Lambda t}(\|\varrho_0\|_{X}^2+\|(\nabla {u}_0, \Delta {u}_0,\partial_i N_0)\|_{L^2}^2),\quad
X=L^2\mbox{  or }H^1,\\[1mm]
&& \| u(t)\|_{H^1 }^2+\| u_t(t)\|^2_{L^2 }+ \int_0^t\|\nabla u(\tau)\|^2_{L^2}\mm{d}\tau\leq
C_\mu e^{2\Lambda t}\|(\varrho_0,\nabla {u}_0, \Delta {u}_0,\partial_i N_0)\|_{L^2}^2,\\[1mm]
&& \| N(t)\|_{L^2}\leq C_\mu e^{\Lambda t}\|(\varrho_0,\nabla {u}_0, \Delta {u}_0,N_0,\partial_i N_0)\|_{L^2}^2,
\end{eqnarray*}
where the constant $C_{\mu}$ only depends  on $\mu$  and $\Lambda$.

(ii) Let $(\varrho, {u}, N)$ solve the linearized Parker problem \eqref{c0104}--\eqref{lincom}.
Then for any $t\geq 0$,
\begin{align}
&\label{uestimatec} \| u(t)\|_{H^1 }^2+\| u_t(t)\|^2_{L^2 } + \int_0^t\|\nabla u(\tau)\|^2_{L^2}\mm{d}\tau\leq
C_{\mu,\mu_0}e^{2\Lambda t}( \| u_0\|_{H^1}^2+I_0), \\[1mm]
&\label{nn0313nc}\| (\varrho,N)(t)\|_{L^2}^2\leq C_{\mu,\mu_0}e^{2\Lambda t}(\|(\varrho_0,N_0)\|_{L^2}^2+\| u_0\|_{H^1}^2+I_0),
\end{align}
where the constant $C_{\mu,\mu_0}$ only depends  on $\mu$, $\mu_0$ and $\Lambda$.
\end{pro}\begin{pf} We prove only the second assertion and the first assertion can be shown in the same manner.
Let $(\varrho,u,N )$ be a solution of \eqref{c0104}--\eqref{lincom}, then $(\varrho, u,N)$ satisfies the identity \eqref{definforI0}.
In view of \eqref{0111nn}, we have
$${E}_{\mm{c}}(u)\leq  \Lambda^2  J(u)  + \Lambda \int(\mu|\nabla\tilde{u}|^2+\mu_0|\mm{div}\tilde{u}|^2)\mm{d} x,  $$
which, combined with \eqref{definforI0}, results in
\begin{equation}\label{0511}
\begin{aligned}
&\int\bar{\rho}|u_t|^2\mathrm{d} x +  2\int_0^t\int (\mu|\nabla u_\tau|^2 +\mu_0|\mm{div} u_\tau|^2)\mm{d} x\mm{d}\tau\\
& \leq I_0+{\Lambda^2}\int \bar{\rho}| {u}|^2\mm{d} x+ \Lambda \int (\mu|\nabla\tilde{u}|^2+\mu_0|\mm{div}\tilde{u}|^2)\mm{d} x.
\end{aligned}\end{equation}

Using Newton-Leibniz's formula and Cauchy-Schwarz's inequality, we find that
  \begin{equation}\begin{aligned}\label{0316}
& \Lambda (\mu\|\nabla u (t)\|_{L^2}^2+\mu_0\|\mm{div} u (t)\|^2_{L^2})  \\
& =K_0+ 2\Lambda\int_0^t\int_{\Omega}\left(\mu\sum_{1\leq i,j\leq 3}\partial_{x_i}\partial_\tau u_{j}
\partial_{x_i}\partial_\tau u_{j} \mm{d} x\mathrm{d}\tau +\mu_0\mm{div} u_\tau \mm{div} u \right)\mm{d} x\mathrm{d}\tau \\
& \leq K_0+\int_0^t(\mu\|\nabla u_\tau \|_{L^2}^2 +\mu_0\|\mm{div} u_\tau \|^2_{L^2})
\mathrm{d}\tau +\Lambda^2\int_0^t(\mu\|\nabla u(\tau) \|_{L^2}^2+\mu_0\|\mm{div} u(\tau)\|^2_{L^2})\mathrm{d}\tau ,
\end{aligned}\end{equation}
where $K_0:=\Lambda (\mu\|\nabla u_0\|_{L^2}^2+\mu_0\|\mm{div} u_0\|^2_{L^2})$.
Thus, we infer by \eqref{0511}--\eqref{0316} that
\begin{equation}\label{inequalemee}\begin{aligned}
&\frac{1}{\Lambda}\|\sqrt{\bar{\rho}}  u_t (t)\|^2_{L^2}+
{\mu}\|\nabla u (t)\|_{L^2}^2 +\mu_0\|\mm{div} u  (t)\|^2_{L^2}\\
& \leq   {\Lambda} \|\sqrt{\bar{\rho}} u (t)\|^2_{L^2}+2 {\Lambda}\int_0^t({\mu}\|
\nabla u(\tau) \|_{L^2}^2+ {\mu}_0\|\mm{div} u (\tau)\|^2_{L^2})\mm{d}\tau +\frac{I_0+2K_0}{\Lambda}.
\end{aligned}\end{equation}
Recalling that
\begin{equation*}\begin{aligned}
\Lambda\frac{\mm{d}}{\mm{d}t}\|\sqrt{\bar{\rho}} u (t)\|^2_{L^2}=2\Lambda\int
\bar{\rho} u (t)\cdot  u_t(t)\mm{d} x\leq\|\sqrt{\bar{\rho}}  u_t (t)\|^2_{L^2}
+\Lambda^2\|\sqrt{\bar{\rho}} u(t)\|^2_{L^2},
\end{aligned}\end{equation*}
 we further deduce from \eqref{inequalemee} the differential inequality:
\begin{equation*}
\begin{aligned}
& \frac{\mm{d}}{\mm{d}t}\|\sqrt{\bar{\rho}} u (t)\|^2_{L^2}+ \mu\| \nabla u (t)\|_{L^2}^2+
\mu_0\| \mm{div} u (t)\|^2_{L^2}  \\
&\leq2\Lambda\left[ \|\sqrt{\bar{\rho}} u (t)\|^2_{L^2} +\int_0^t(\mu\| \nabla u (\tau)\|_{L^2}^2 +
{\mu}_0\|\mm{div} u(\tau) \|_{L^2}^2) \mathrm{d}\tau\right]   +\frac{I_0+2K_0}{\Lambda}.
\end{aligned}
\end{equation*}

 Applying Gronwall's inequality \cite[Lemma 1.2]{NASII04} to the above inequality, one concludes
 \begin{equation}\label{estimerrvelcoity}
\begin{aligned}
  \|\sqrt{\bar{\rho}} u (t)\|^2_{L^2}+\int_0^t({\mu}\|\nabla u (\tau)\|^2_{L^2}+{\mu}_0\|\mm{div} u(\tau) \|^2_{L^2})\mm{d}\tau
\leq\left[\|\sqrt{\bar{\rho}}u_0\|_{L^2}^2+\frac{(I_0+2K_0)}{2\Lambda^{2}}\right] e^{2\Lambda t},
 \end{aligned}  \end{equation}
which, together with \eqref{inequalemee}, yields
\begin{equation*}\label{uestimate1n}\begin{aligned}&
\frac{1}{\Lambda}\|\sqrt{\bar{\rho}}  u_t (t)\|^2_{L^2} +{\mu}\|\nabla u (t)\|_{L^2}^2 +\mu_0\|\mm{div} u  (t)\|^2_{L^2}\\
& \leq 2\left[\Lambda\|\sqrt{\bar{\rho}}u_0\|_{L^2}^2+\frac{(I_0+2K_0)}{2\Lambda }\right] e^{2\Lambda t} +\frac{I_0+2K_0}{\Lambda}.
\end{aligned}    \end{equation*}
Thus \eqref{uestimatec} follows from the two estimates above.
Finally, using \eqref{lincom}$_1$, \eqref{lincom}$_2$  and \eqref{estimerrvelcoity}, we find that
\begin{equation*}\begin{aligned}
\|(\varrho,N)(t)\|_{L^2}\leq & \|(\varrho_0,N_0)\|_{L^2}+\int_0^t\|(\varrho,N)_\tau \|_{L^2}\mm{d}\tau \\
\leq &\|(\varrho_0,N_0)\|_{L^2}+C\int_0^t\| u(\tau)\|_{H^1}\mm{d}\tau \\
\leq & C_{\mu,\mu_0}e^{\Lambda t}(\|(\varrho_0,N_0)\|_{L^2}+\| u_0\|_{H^1}+\sqrt{I_0}),
\end{aligned}\end{equation*}
and get \eqref{nn0313nc}.
The proof is complete.
  \hfill $\Box$
\end{pf}
\vspace{4mm} \noindent\textbf{Acknowledgements.}
The research of Fei Jiang was supported by NSFC (Grant Nos. 11271051,
11301083 and 11471134), and the research of Song Jiang by the National Basic Research Program
under the Grant 2011CB309705 and NSFC (Grant Nos. 11229101 and 11371065).
\renewcommand\refname{References}
\renewenvironment{thebibliography}[1]{%
\section*{\refname}
\list{{\arabic{enumi}}}{\def\makelabel##1{\hss{##1}}\topsep=0mm
\parsep=0mm
\partopsep=0mm\itemsep=0mm
\labelsep=1ex\itemindent=0mm
\settowidth\labelwidth{\small[#1]}%
\leftmargin\labelwidth \advance\leftmargin\labelsep
\advance\leftmargin -\itemindent
\usecounter{enumi}}\small
\def\newblock{\ }
\sloppy\clubpenalty4000\widowpenalty4000
\sfcode`\.=1000\relax}{\endlist}
\bibliographystyle{model1b-num-names}

\end{document}